\documentstyle[11pt,aaspp4]{article} 
\newcommand{\kms}{km~s$^{-1}$}
\newcommand{\kmsM} {km~s$^{-1}$~Mpc$^{-1}$}

\newcommand{\etal}{{\it et al.\/}}

\lefthead{Cohen {\it et al.} }
\righthead{HDF Redshift Survey}

\begin{document}

{\bf May 24, 2000}

\vspace{10 mm}

{\bf On December 2, 1999, I archived in the LANL server a preprint (Astro-ph/9912048)
of the paper
``Caltech Faint Galaxy Redshift Survey X: A Redshift Survey in the Region
  of the Hubble Deep Field North'' by
Judith Cohen, David Hogg, Roger Blandford, Lennos Cowie, Esther Hu,
  Antoinette Songaila, Patrick Shopbell and Kevin Richberg.  This paper
is in press in the ApJ.  The paper as submitted to Astro-ph at that time
did not contain certain key figures and tables.

The text of this paper has today been replaced with a version which contains all
the figures and tables, including the galaxy redshifts.

Unfortunately, I
and my group, being small (because I have received no federal funding for
this research for several years) have not finished two 
final pieces of analysis: (1) a paper
on the evolution in the galaxy luminosity function and (2) a paper on the
relationships between broadband colors, morphologies, and narrow spectral
features (lines and breaks).  I therefore request that the redshifts not
be used for projects which are substantially similar to either of these
two projects until I have submitted such papers to astro-ph.  I expect
this to happen by 2000 October.  Thank you very much.}

\vspace{10 mm}
\bf{Judy Cohen}

\clearpage

\title{Caltech Faint Galaxy Redshift Survey X: 
A Redshift Survey in the Region of the Hubble Deep Field North\altaffilmark{1}}

\author{Judith G. Cohen\altaffilmark{2}, David W. Hogg\altaffilmark{3,4,5},
Roger Blandford\altaffilmark{3}, 
Lennox L. Cowie\altaffilmark{6}, Esther Hu\altaffilmark{6},
Antoinette Songaila\altaffilmark{6}, Patrick Shopbell\altaffilmark{2,7}
\& Kevin Richberg\altaffilmark{2} }

\altaffiltext{1}{Based in large part on observations obtained at the
	W.M. Keck Observatory, which is operated jointly by the California 
	Institute of Technology and the University of California}
\altaffiltext{2}{Palomar Observatory, Mail Stop 105-24,
	California Institute of Technology, Pasadena, CA \, 91125}
\altaffiltext{3}{Theoretical Astrophysics, California Institute of Technology,
	Mail Stop 130-33, Pasadena, CA \, 91125}
\altaffiltext{4}{Current Address:  Institute for Advanced Study, Olden Lane, Princeton, NJ \, 08540}
\altaffiltext{5}{Hubble Fellow}
\altaffiltext{6}{Institute for Astronomy, University of Hawaii, 2680
Woodlawn Drive, Honolulu, Hawaii 96822}
\altaffiltext{7}{Current Address: Department of Astronomy, University of Maryland, 
       College Park, MD \,20742-2421}

\begin{abstract}

A redshift survey has been carried out in the region of the Hubble
Deep Field North using the Low Resolution Imaging
Spectrograph at the Keck Observatory.  The resulting 
redshift catalog, which contains 671 entries, 
is a compendium of our own data together with published LRIS/Keck data.  
It is more than 92\% complete for objects, irrespective of morphology,
to $R = 24$ mag
in the HDF itself and to $R = 23$ mag in the Flanking Fields within
a diameter of 8 arcmin centered on the HDF, an unusually high completion
for a magnitude limited survey performed with a large telescope.
A median redshift
$z = 1.0$ is reached at $R \sim 23.8$. 

Strong peaks in the redshift distribution, which arise when a group
or poor cluster
of galaxies intersect the area surveyed, can be identified to $z \sim 1.2$
in this dataset.  More than 68\% of the galaxies are members 
of these redshift peaks.  In a few cases, closely spaced peaks in $z$
can be resolved into separate groups of galaxies that can be distinguished 
in both velocity and location on the sky.  

The radial separation of these peaks in the pencil-beam survey
is consistent with a characteristic length scale for the their separation
of $\approx$70 Mpc in our adopted cosmology ($h = 0.6, \Omega_M = 0.3$, $\Lambda = 0$).
Strong galaxy clustering is in evidence at all epochs back to $z \le 1.1$.

A near-infrared selected sample with $K < 20$ was also constructed in this field.
Extremely red objects with $R - K > 5.0$ comprise 7\% of the total 
$K$-selected sample.
This fraction rises rapidly towards fainter $K$ mag, reaching about 10\% at
$K \sim 19.7$. 

We have attempted to identify the radio sources in the region of the HDF.
The secure radio sources seem to divide into two classes.  The first have 
reasonably bright galaxies at moderate redshifts as
optical counterparts, while the second,
comprising about 1/3 of the total,
have extremely faint optical counterparts ($R \ge 25$ mag).  These
do not represent a continuous extrapolation in any property
($z$ or dust content) of the first group.

We identify $\sim$2/3 of the secure mid-IR sources in the region of the HDF
with normal galaxies
with $z < 1.3$.  The ratio of the mid-IR to optical flux increases as
$z$ increases, but this is due primarily to selection
effects, and the same trend is seen in the radio sources.  
We suggest that the mid-IR emission is more tightly coupled to the rate of
ongoing star formation than is the radio emission.

We also demonstrate that the best photometric redshift techniques are 
capable of reaching
a precision of $\sigma[(z_{phot} - z_{spec})/(1+z_{spec})] = 0.05$ for 
the majority of galaxies with $z < 1.3$.

The two broad lined AGNs with $z < 3$ are the brightest objects
in the redshift peak at $z \sim 0.96$.

\end{abstract}


\section{Introduction}

The Hubble Deep Field North (henceforth HDF) is a region of intense astrophysical interest.
The selection of the field and the
HST campaign in the HDF are described by Williams \etal\ (1996).
The field was chosen to be a ``typical'' high galactic latitude field
with no known bright sources in the x-ray, UV, optical, IR and radio
passbands, and no known nearby ($z < 0.3$) galaxy clusters. 
The superb multi-color WFPC2 images taken by the Hubble Space Telescope
reach to $B \sim 29$ and with their 0.1 arcsec spatial resolution
provide an unprecedented view of the distant Universe.

There has been follow up of these beautiful HST images of the
HDF in most wavelength ranges. These include the deep radio survey of Richards \etal\ (1998) 
with the VLA, which extends into the Flanking Fields 
around the HDF, and the survey by Hughes \etal\ (1998) 
in the sub-mm.  Space-based IR followup with NICMOS
was done by Thompson, Storrie-Lombardi \& Weymann (1999) and by Dickinson (1999), 
while Aussel \etal\ (1999) report on a
survey of the HDF with ISO.  From the ground,
Hogg \etal\ (1997) imaged two subregions to $K = 25.2$
at 2.2 $\mu$. 

The strong interest in this field has also led to major efforts by
several groups to obtain redshifts for the galaxies in this field to
study star formation, galaxy evolution and galaxy clustering,
as well as to aid in the interpretation of data at other wavelength regimes
where redshifts are more difficult/impossible to determine.
Given the magnitude limit for practical redshift determinations, extensions
beyond that regime using photometric redshifts have been actively pursued
by various groups and applied to the HDF to explore star formation rate
at very high $z$, e.g. Connolly \etal\ (1997).

We provide here
a compilation of our own work over the past few years
towards a redshift survey in the
HDF and its environs with published work to produce a 
very complete magnitude limited redshift survey of objects in this region.
All spectra, both our own and from the literature,
were taken with the Low Resolution Imaging 
Spectrograph (henceforth LRIS) (Oke \etal\ 1995) at the Keck Observatory,
without which this effort would be impossible.

We present the redshift catalog in \S\ref{redcat}, which greatly extends our
earlier work in the region of the HDF (Cohen \etal\ 1996, henceforth C96)
from 140 objects to 671 objects with redshifts.
After a brief
discussion of the extremely red objects in \S\ref{eros}, we comment on the accuracy
of photometric redshift techniques in \S\ref{photoz}.  We then proceed
in \S\ref{zdist} to discuss the distribution in redshift of the sample,
the nature of the strong redshift peaks and large scale structure.
A sketch of the
luminosity and density evolution of galaxies in the region
of the HDF is given in \S\ref{lumevol}.
We follow with a discussion of the HDF as seen in other wavelength regimes
(radio, mid-IR and sub-mm) in
\S\ref{radio}, and then summarize our results in \S\ref{summary}.

We adopt the cosmological parameters $H_0 = 60$~\kmsM\ and 
$\Omega_M = 0.3$ with $\Lambda = 0$, as in earlier papers in this series.

\section{The Redshift Catalog\label{redcat}}

\subsection{Observing Procedure}

The goal established by the Caltech group was to carry out
a magnitude limited survey of objects in the
region of the HDF to a limit of $R = 24$ for the HDF itself
\footnote{The HDF is defined as the area on the sky deeply imaged by the HST,
excluding the PC detector, as discussed in Williams \etal\ (1996).}
and to a limit of $R = 23$ mag
for the region within a field with a diameter of 8 arcmin
centered on the HDF, but excluding the HDF itself, which we call the
Flanking Fields. 
\footnote{The center of the circle is at RA = 12 36 50.00, Dec = +62 12 55.0. (J2000)} 
The size of the region we included in the spectroscopic survey 
was determined by the
length of the LRIS slit in the multislit mode. 
Hogg \etal\ (1999) (henceforth H99)
present a four-color photometric catalog
for the region of the HDF covering this area; their $R$ catalog is adopted here
to define the magnitude limited samples.
\footnote{The calibration of the photometry is described in H99.  The
zero point is Vega-relative.  The $K$ photometry uses a $K$-short filter.
The catalog seems to be secure to $R$ = 24 and contains entries
to $R \sim 25$.}
Objects were included in the sample independently of their morphology
although, of course, there may be extremely extended, low surface brightness
sources that would not be detected.

Cowie, Hu \& Songaila (1995) in general pursued more targeted
projects, and include
the majority of their spectra in the region of the HDF in the present
compilation.

Observations were carried out and analyzed in a manner
similar to that described in Cohen \etal\ (1999b) from
1996 to 1999 using many slit masks in the region of the HDF.  The
gratings used and the resulting spectral resolution and spectral coverage are
not uniform throughout the catalog, although most of the spectra
were obtained with the 300 g/mm grating, giving a spectral resolution
with a 1 arcsec slit of 10 \AA.  Spectra of some of the fainter
objects were obtained with a 155 g/mm grating which became available in
1998; the sky subtraction is more difficult at lower dispersion
and most of the objects with dubious features redward of 
6000 \AA\ were subsequently re-observed
with a higher dispersion grating.  A few objects were observed 
with the higher dispersion 600 g/mm gratings.  A typical 
uncertainty in redshift is 200 \kms\ in the observed frame for an object with a
spectrum taken using the 300 g/mm grating and with $R < 23$.

\subsection{Sources of Published Data}

The sources of published data at lower redshifts ($z \lesssim 1.3$)
are Phillips \etal\ (1997) (from the Lick Deep Group) and the
work of the Caltech and of the Hawaii group published in C96.
At high redshift ($z > 2)$, the results of
Steidel (Steidel \etal\ 1996,
Dickinson 1998, Steidel \etal\ 1999) and of the 
Lick Deep Group (Lowenthal \etal\ 1997) are combined
with single very high redshift galaxies found by Spinrad \etal\ (1998),
Weymann \etal\ (1998) and Waddington \etal\ (1999).
Five unpublished
redshifts from C. Steidel (private communication) are also included. 

As the catalog was compiled, any discrepancies between multiple 
observations of the same object were reconciled. In most cases 
with duplicate
observations, the agreement was superb ($\Delta(z) \le 0.005$).  For
about 20 galaxies with more than one spectrum, 
the initial redshifts were wildly discrepant, but these cases were
resolved by one person (JGC and sometimes also LLC) looking at all
the available spectra of the galaxy.
Details for the small number of galaxies for which the final redshift
adopted in this catalog is different from the previously
published value can be found
in the Appendix.

Table~1 lists the sources of the data and the number of objects
included from each.

\subsection{Matching the Photometric and Redshift Catalogs}

The coordinate system in the HDF itself, defined now by VLA
observations, has changed very slightly with time
as described in Williams \etal\ (1996), while the definition of
the coordinate system in the Flanking Fields is described in H99.
As will be shown in \S\ref{radiosrc}, the coordinate system
we use agrees with that defined by the VLA to within 0.1 arcsec in RA
and also in Dec.
The redshift catalog was constructed beginning with coordinates 
defined by the observer for each object.  
Then the small systematic offsets in RA and Dec between each 
set of redshift data were determined, and the appropriate (small)
corrections were made to the object coordinates.

The matching of the redshift catalog 
onto the photometric catalog of H99 was non-trivial,
particularly in the HDF, where the depth of the sample increases
the crowding,
and particularly for objects with $R > 23$.  
The matching was done automatically for objects with $R < 23.5$ mag;
if only one object in the photometric catalog was within 1.5 arcsec 
of the position of an object in the redshift catalog, it was
accepted as the match.  

Objects fainter than $R = 23.5$ as well as brighter
ones with multiple or no matches in the photometric catalog of H99
were examined individually.  In most cases, the match was obvious.
In all but a few of the remaining cases, the nature of the problem
became apparent after checking for close pairs,
coordinate errors, bookkeeping errors, etc.
There are a very small number of objects with uncertain matches,
and these are noted in the redshift catalog.

There are some objects in the region of the HDF
for which redshifts have been obtained
which are genuinely too faint for the photometric catalog of H99.  These
have been included as a supplemental table in the photometric
catalog (see H99 for details).  They are indicated by notes in Table 2.

\subsection{Close Pairs}

The SExtractor algorithm (Bertin \& Arnouts 1996) with the parameters 
adopted by H99 in some cases
fails to separate close pairs with separations of under 1.5 arcsec.
When there are two entries in the redshift table, but a single corresponding
entry in the photometric catalog, photometry of the two components
can be found in a supplemental table in H99 and there is an appropriate
note in Table 2.  For purposes of
computing completeness, we assume that the single entry 
representing the pair in the
primary photometric table has been matched only if both members of the
pair have redshifts.

A discussion of the objects that are genuine close pairs rather than
projections and the implications for merger rates is given in
Carlberg \etal\ (2000).

\subsection{Completeness}

In the HDF itself only a few faint objects
near the magnitude cutoff are missing
from the catalog.  In the Flanking Fields, however,
some brighter objects were missed initially due to problems in defining the sample.
Since the final photometric catalog only became available very late in the project,
it was not possible to pick up all the 
missing objects in the limited telescope time available.  
Hence some of the missing objects in the Flanking Fields
are not near the faint limit, and spectroscopic observations,
had they been carried out,
would undoubtedly have yielded a definite redshift.

The redshift completeness fraction is very high,
as is shown in Figure 1a,b, which also shows the cumulative redshift 
completeness for objects
brighter than a given $R$ mag.
Of the 114 objects with $R < 24$ in the $R$ band photometric catalog
of H99 in the HDF, only 8, the
brightest of which has $R = 23.2$,
are missing from the redshift catalog.  The completeness is thus 93\%.
In the Flanking Fields within an area with a diameter of 8 arcmin 
centered on the HDF,
and excluding the HDF itself, there are 434 objects in the $R < 23$
photometric sample, and the redshift completeness there also 
exceeds 92\%.

The eight objects in the HDF in the $R$-selected sample ($R < 24$)
that do not have spectroscopic redshifts at the present time and
hence
that are not included in Table 2b are listed in an appendix.
Many of these have been observed more than once.
It should be empahsized that the objects in the region of the HDF
that were observed, but for which redshifts have not been determined 
yet, are not included in the tables nor considered when 
calculating completeness.

\subsection{Galaxy Spectral Classes and Redshift Quality Classes}

The galaxy spectral classification scheme adopted here is that of
Cohen \etal\ (1999b).  To review briefly,
``${\cal E}$'' galaxies have spectra dominated by emission lines,
``${\cal A}$'' galaxies have spectra dominated by absorption lines,
while  ``${\cal I}$'' galaxies are of intermediate type.   Galaxies
with broad emission lines are denoted as spectral class ``${\cal Q}$''.
Starburst galaxies showing the higher Balmer lines 
(H$\gamma$. H$\delta$, etc.) in emission 
are denoted by ``${\cal B}$'', but for such faint
objects, it was not always possible to distinguish them from 
``${\cal E}$'' galaxies.

All the spectral classifications were done by JGC. 

The spectral
classifications are most accurate for the brighter galaxies with
$z < 1.2$.  For the fainter ones, emission lines are easier to detect
than absorption lines, and the distinction between galaxy spectral classes
``${\cal E}$'', ``${\cal EI}$'' and ``${\cal I}$'', which depends
critically on the presence or absence of the Balmer lines in absorption
near 4000\AA\ in the rest frame, becomes blurred.
In addition, for some galaxies with low redshifts,
the spectral region covered did not extend sufficiently to the
blue, and the 4000 \AA\ region was lost, so the distinction
between ``${\cal E}$'', ``${\cal EI}$'' and ``${\cal I}$'' could not be
made with the material available.  Notes to the table indicate
such concerns.

Spectral classes for galaxies with $z > 1$ were not assigned in a 
completely consistent manner.  A galaxy whose spectrum 
showed emission at 3727 \AA\ and absorption at the 2800 \AA\ Mg doublet,
a fairly typical combination at $z \sim 1.2$,
could be assigned to spectral class ``${\cal E}$'', ``${\cal EI}$'' or ``${\cal EA}$''.

The spectra of galaxies with $z > 2$ could not be classified as carefully.
Most of these are from published data, and in some cases the original
spectra were not available to JGC.  In such cases (and also for
lower $z$ objects where the spectra were not available to JGC), the spectral
class is given as ``${\cal X}$''.
(Unless otherwise noted, when a spectral type is required, for 
example to define the symbol
in a figure, these are assumed to be ``${\cal E}$'' galaxies.)

Redshift quality classes were assigned following Cohen \etal\ (1999b).
One additional class was added here for 
extremely faint objects ($R > 24.3$)
observed by the Caltech group which turned up serendipitously in slitlets intended
for some brighter nearby galaxy and which show only a single emission
line with no sign of a continuum.  In our previous work, 
for spectra showing only a single emission line,
the assumption was always made that this line is the [OII] line at 3727 \AA.
However, in these particular cases, the
objects are so faint that we have instead assumed that the emission
line is Ly{$\alpha$}.  There are seven galaxies in this category assigned
to redshift quality class 11.  These are probably similar to the $z \sim 4$
single emission line galaxies found by Hu, Cowie \& McMahon (1998).  
There is one additional
galaxy which would have been in this category except that a second
spectrum is available from Lowenthal \etal\ (1997)
which confirms the $z \sim 3$ redshift.

No redshift quality class is assigned for the ``${\cal X}$'' galaxies.

\subsection{The Merged Redshift Survey}

Table 2a,b contains the merged redshift table.  The objects are in order of
RA, with those in the the Flanking Fields in Table 2a, and those in
the HDF itself in Table 2b.\footnote{It is believed that the boundary of the
six sided polygon defining the HDF used here
coincides with that of Williams \etal\ (1996) to within 1 arcsec.
However, for sources very close to the boundary of the HDF, both Table
2a and Table 2b should be checked.}  The coordinates listed are those
of H99, from the matching of the redshift observations to the $R$-band 
object catalog.  The $R$ magnitude is that of the matched object from H99 as well.
Then follows the redshift, the redshift
quality class, the galaxy spectral class and the source(s) for the redshift determination.

Following the main catalog, Table 3 contains similar data for 15
objects just outside the survey boundary (a diameter of 8 arcmin from the
center of the HDF).

There are 671 entries.  Of these,
146 are within the three WF chips of the HDF itself, 510 are in the Flanking
Fields within a diameter of 8 arcmin centered on the HDF, but outside the HDF
itself, and 15 are just outside the spatial boundary of the survey.
There are 11 spectroscopically confirmed galactic stars in the HDF and 
42 spectroscopically confirmed stars in the Flanking Fields.

Figure~2 displays the HST image of the HDF 
from Williams \etal\ (1996) with the objects and their redshifts marked.
Figure~3 is a similar overlay of four panels
identifying the objects in the redshift survey on each of the four quadrants of the 
the HST composite image of the Flanking Fields.  A color
coding of the galaxy spectral class is used for objects with $z < 1.0$.

\subsection{Median Redshift}

The median redshift as a function of $R$ and of $K$ is shown
in Figure~4 and given in Table~4. The first and last quartiles are also
tabulated for $R$.  Although redshifts have
been determined for more than 92\% of the sample to its
defined magnitude limits, 
at the fainter magnitudes these are biased by the
incompleteness of our spectroscopic survey, in particular, by the
detection efficiency, or lack thereof, in the regime $1.4 < z < 2.0$.

We have corrected the sample in an approximate manner by adding in 
a rough representation of the set of missing galaxies in the interval
$1.4  < z < 2.0$ derived from the magnitude distribution of the sample
near $z \sim 1$.   Specifically for the missing galaxies, we assumed
the luminosity function to be the same
as that in the interval $1.0 < z < 1.2$
and the comoving density
to be constant with $z$ in this regime.
The augmented set of galaxies
then has a redshift distribution whose
medians and quartiles are listed in table 4 as ``corrected'' values.

Our redshift catalog for the region of the HDF
becomes highly incomplete at $R > 24$, and it is unclear
how many low $z$ objects should be added at that magnitude level that
were rejected by the UV-dropout selection technique used to select
samples for spectroscopy by various groups. 

Table~4 shows that a median $z$ of 1.0 is reached at $R = 24.2$ among the observed
sample in the region of the HDF.  This is
reduced to $R = 23.8$ when
one corrects as described above for the galaxies in the regime $1.4  < z < 2.0$
whose redshifts were probably not determined in our survey.

\subsection{Extremely Red Objects\label{eros}}

We adopt the criterion that objects with $R - K > 5.0$ are considered
extremely red objects (EROs).  The $K-$selected sample 
from H99 in the region of the HDF is complete for $K \lesssim 19.8$
and, as is shown in Figure~4 of H99, 
covers an area only slightly larger than that of the spectroscopic
sample.
There are 33 EROs out of 487 objects in the photometric catalog,
which is 7\% of the total $K-$selected sample,
or 8\% of the total sample excluding galactic stars.  The fraction of
EROs as a function of $K$ rises rapidly from 4\% of the objects 
at $K \sim 18.3$ 
to 10\% at $K \sim 19.7$.   The reddest ERO has $R - K = 6.0$.
Only one of these EROs in the region of the HDF, 
H36443\_1133, has a redshift.  Its spectral class is, as expected,
``${\cal A}$'' with $z = 1.05$.  

There is an additional ERO in this sample with a redshift, F37016\_1146,
with $z = 0.884$, which is a radio source and is fainter than the limit
of the H99 photometric survey in $R$.  This galaxy shows 3727\AA\
in emission and is discussed in Barger, Cowie \& Richards (2000).

The nature of the ERO population is still uncertain.
Some, particularly the most luminous and reddest ones, 
may be distant dusty
mergers or starbursts, but we proposed in
Cohen \etal\ (1999a) that
the majority of the EROs in this color and magnitude range
are old galaxies at $z \sim 1.5$.
Although EROs are very rare in optically selected samples,  their presence
in substantial numbers in faint infrared selected samples
such as the study of NICMOS images by
Benitez \etal\ (1999), as well as our work and that of others, 
is by now well established, in contrast to the claims of
Kauffmann \& Charlot (1998).
Their existence forces a reconsideration
of models for the formation of
elliptical galaxies in which the bulk of such galaxies form
at $z \sim 1$ (e.g. that of Kauffmann, Charlot \& White 1996).

\section {A Blind Check of Photometric Redshifts\label{photoz}}

We have utilized our new merged catalog of redshifts for objects in the
HDF from Table 2b
to carry out a second blind test of photometric redshifts in a spirit 
similar to our first test, Hogg \etal\ (1998).
The advantage of the present test is that the set of spectroscopic redshifts
known to us but not available to the developers of photometric redshift
algorithms is
significantly larger than it was at the time of our first test.
In addition, the redshifts used in Hogg \etal\ (1998) were preliminary values,
while those of Table~2 are final redshifts.

Three photometric redshift schemes were chosen to span the
range of techniques used.  Wang, Bahcall \& Turner (1998) use polynomial fits
to colors, while Sawicki, Lin \& Yee (1997) and 
Lanzetta, Yahil \& Fernandez-Soto (1996) 
use template spectra of galaxies.  The catalogs of predicted photometric redshifts
were taken from each group's WWW site as listed in their papers.

In each case, the list of redshifts for the training set of galaxies 
was checked against the entries in Table 2b.  If they agreed to within the larger of
5\% or 0.05, the object was considered a good calibrator.  If they
disagreed (often because the preliminary redshift adopted by Hogg \etal\ 1998
was subsequently substantially modified)
the object was considered as an unknown.  To this we
add the galaxies in the HDF whose redshifts, even in preliminary form,
were not available for each of the photometric redshift training sets.  This
is a number which varies depending on when the photometric redshift
technique was developed and applied to the HDF, as more calibrating 
galaxies with spectroscopic
redshifts became available with time.

We compute the mean and dispersion  of the difference between
the predicted redshifts
from the photometric technique and the measured spectroscopic
redshift for this set of unknowns for each of the three groups,
as well as the correlation coefficient.
These are calculated in units of (1+z) to avoid errors in small redshifts
artificially inflating the results.
In each case, H36396\_1230 (a probable AGN) was omitted;
none of the groups came close to predicting this object's redshift.

Figure~5 illustrates the results for the three techniques
in the low $z$ regime ($z < 1.5$).   A distinction is made between 
spectroscopic redshifts that are secure and those that are more uncertain
(i.e. have quality class 3, 8 or 9).
There are a few outliers, some of which are beyond
the range displayed, but most of
the galaxies show a small dispersion in $(1+z)$ between the
predicted redshift and the measured spectroscopic value.  

Table~5 gives the results for each of the three groups, 
tabulated separately for $z< 1.5$  and 
for $z > 1.9$.   The first column contains the total number of galaxies with new spectroscopic
redshifts, followed by the mean and 1$\sigma$ dispersion of $\Delta(1+z)$.
This is repeated after eliminating the outliers at more than 4$\sigma$.

At low $z$, once the outliers comprising ${\sim}10$\% of the objects
are eliminated, both the astrophysically motivated scheme of Sawicki \etal\ (1997)
and the polynomial fitting of Wang, Bahcall \& Turner (1998) give 
remarkably good results.  At the time that
Lanzetta \etal\ (1996) carried out their work, fewer spectroscopic
redshifts were available to use as calibrators,
which may explain the somewhat larger dispersion between their photometric
redshifts and the recent spectroscopic measurements.

At high $z$, the astrophysically
motivated schemes work somewhat better than the polynomial fitting, but all perform
surprisingly well.

There are four galaxies that are rejected as outliers by all of the
photometric redshift techniques tested here.
\footnote{The four galaxies are H36384\_1234, H36414\_1142, H36493\_1317
and H36569\_1302.}
Three of these galaxies have uncertain redshifts and one of the three
is 1.7 arcsec away from a much brighter galaxy, 
which may affect its photometry.  The fourth outlier has a secure redshift.
However, the fraction of outliers is small, the success rate is very high, 
and the precision of the predictions is extremely good.

This, our second blind test, has again demonstrated that photometric
redshifts with suitable algorithms are capable of producing
highly accurate predictions for galaxy redshifts over the magnitude
and redshift intervals for which they have been calibrated.  
The best of the three schemes tested
predicts (1+z) to less than  5\% for more than 
90\% of a realistic sample of galaxies over the full range of
$z$ for which training galaxies exist.  A dispersion of only 15\%,
still remarkably small,
covers the maximum found among the three schemes.

The current lack
of training sets of galaxies
with spectroscopically determined redshifts in the regime 
$1.2 < z < 2.0$ is a problem which should be resolved shortly
through the efforts of several groups, including our own.
Photometric redshift techniques are becoming useful tools for
many problems,  including determining luminosity functions and
star formation rates, choosing
samples for spectroscopy, and related issues.  It is of course
a requirement that suitable multi-color high precision photometry 
exist and that the 
spatial resolution available is sufficient to avoid
image overlapping within the magnitude regime of interest.

\section{The Distribution in Redshift\label{zdist}}

\subsection{The Strong Redshift Peaks}

The redshift distribution from our catalog in the region of the HDF
is shown in Figures 6-8 as a function of $R$ mag
in progressively increasing level of detail.  The galaxy spectral types
are indicated by the same set of symbols, described in
detail in the caption of Figure~6, in all the figures.
In Figure~6 the gap near $z \sim 1.7$ where it is currently
quite difficult to determine redshifts is apparent.
Figures~6 and 7 give an indication of the total distribution, and show that the 
luminosity of the brightest objects is declining in a way more or less similar
to that expected, a subject to which we return later.

Figure~8a,b presents the redshift distribution at a scale where individual
peaks, apparent in a magnitude-redshift plot as vertical lines,
can be seen.  The $z$
distribution is highly structured, with peaks that
first become visible at $z \sim 0.3$.
(At lower redshift the sample is too sparse because the cosmological volume
included is too small and because the HDF was selected to be devoid of
bright galaxies.) In the region of the HDF, we can follow
these structures out to $z \sim 1.1$ without difficulty
because of the larger sample.

The redshift peaks were found using the Gaussian kernel algorithm
described in Cohen \etal\ (1999a).  The calculation is carried out
in local velocity space $V = c~\rm{ln}(1+z)$ with a smoothing of
15,000 \kms\ to define the overall distribution in redshift and
a velocity width of 300 \kms\ for the smoothed distribution.
The overdensity is shown as a function of $V$ in Figure~9, where
the rich suite of groups present in the data is readily apparent.
Structures with a maximum overdensity of 3 
or larger are accepted as real discrete groups
of galaxies and are
listed as the statistically significant sample of redshift peaks
in Table~6.  There were two additional probable structures picked
out by eye at $z = 0.638$ and $z =0.903$ which fall below
the adopted overdensity threshold.
All 6 redshift peaks from 
our preliminary work in the HDF reported in C96
appear in Table~6, but with our much larger and fainter redshift
sample we can see more peaks extending to higher redshift.

The overdensity statistic is a measure of the
size of a group which to first order is independent of $z$, unlike galaxy
counts of members of a group to a fixed limiting magnitude.
It is interesting to note
that (ignoring the regime $z < 0.1$ where the statistics are small and the
background signal is mostly from higher $z$ objects), the peak value reached
by the overdensity
is constant to within a factor of two with redshift over the regime
$0.1 < z < 1.3$.

As in our earlier work, we note that the ``$\cal A$'' galaxies preferentially
reside in these peaks and are among the brightest in each peak 
up to the limit where they can be reliably detected, i.e. $z < 0.8$.

\subsection{The Spatial Extent of the Redshift Peaks}

The field studied here is 8 arcmin in diameter, which corresponds to
3.0 Mpc at $z = 0.5$ and to 3.7 Mpc at $z = 1$.  This means that
we can only probe structure at the length scale of groups and clusters, and
cannot discern transverse structure at any larger scale.  Figures~10 and 11
show the spatial distribution for the highest major redshift peak detected,  
at $z = 1.015$, and for the very populous redshift peak
at $z = 0.478$.  In the
case of the latter, the spatial distribution is highly non-uniform;
there appear to be two groups $\sim$1.5 Mpc apart.  The spatial distribution 
of galaxies in the peak at $z = 1.016$
also appears
non-uniform, with a group/cluster centered about
100 arcsec SW of the center of the HDF.

\subsection{The Velocity Dispersions of the Strong Redshift Peaks}

The velocity dispersions for the sample of statistically significant
redshift peaks are listed in Table~6. These have
been computed using the biweight algorithm of Beers, Flynn \& Gebhardt (1990)
because of its resistance to the presence of outliers.  The mean and
sigma for each structure
are calculated from a sample which in general extends over
the regime $z_p{\pm}0.010$.  
The instrumental contribution to $\sigma(v)$, estimated to be
200 \kms\ in the observed frame, has been removed in quadrature.
The number of galaxies in each peak
is estimated by counting the number within 1$\sigma$ of the peak
and assuming Gaussian statistics.  Adding up the total
estimated membership for the set of statistically significant peaks, we
find that 68\% of the galaxies are in groups with
an maximum overdensity of larger than 3,
in agreement with the results from  
Cohen \etal\ (1999a) and from our earlier work with a much smaller
sample in the HDF (C96).  

%
%
In most cases the peak finding algorithm was successful in
decomposing close peaks into their separate components.
This is evident in the region around $z \sim 0.08$ and $z \sim 0.48$.
However, the region around $z \sim 0.41$ was manually divided
into two separate groups as indicated in the notes to Table~6.
These three cases are believed to represent multiple groups 
within a single redshift peak.
Now that we have a much larger sample covering a larger area on the sky 
it should not be surprising that we find 
more than one group of galaxies within the same redshift peak.

The very low velocity dispersions achieved in some cases (there
are four entries in Table~6 with $\sigma_v < 200$ \kms) is very gratifying
and supports the quoted errors of our redshifts.

The individual groups contributing to the large redshift peak at $z_p = 0.48$ can
be separated not only in velocity space but also
spatially within the area on the sky covered by our survey.  
The three panels of Figure~12 show first
the velocity structure of this redshift regime
at even higher resolution than Figure~8 , then present a decomposition 
of this structure (shown as a whole
in Figure~11) into two separate groups with
$z_g = 0.476$ and 0.485.  These two groups have a transverse
separation of 1.7 Mpc and a difference in mean velocity (in the rest frame) of
350 \kms.  There is a third smaller group that probably can be split off
from the second of these at about the same redshift, $z_g = 0.474$,
with its center close to the center of our field.  They may be in
the process of merging to form a cluster.

Thus the velocity dispersions within the redshift peaks must be treated with caution.
Contributions by multiple groups may artificially inflate their values.

\subsection{The Radial Separation of the Redshift Peaks}

At this point we introduce an additional assumption that the redshift
peaks seen so prominently in our sample represent the intersection
of our pencil beam with ``walls'' resembling those  
discovered locally from the CFA Survey 
by de Lapparent, Geller \& Huchra (1986).
The spatial scale of these ``walls''
as seen locally is very large.   
Doroshkevich \etal\ (1996) finds from the LCRS that most galaxies are in
sheets separated by 77 ${\pm}9 h^{-1}$ Mpc (128 Mpc in our cosmology).
Broadhurst \etal\ (1990) (see also Willmer \etal\ 1994) 
claimed to detect strictly periodic ``walls'' with a spacing of
213 Mpc in our cosmology out to $z \sim 0.2$ from pencil beam surveys.

The only handle we have with our dataset for the region
of the HDF on such large scale structure is to look along the line of sight
at the separation of the redshift peaks.  We must assume that 
for most ``walls'' the  area on
the sky in our survey is sufficiently large that at least one 
group or sparse cluster
will be located within the area surveyed.  

We further assume that the ``tilt'' in redshift of a ``wall'' across 
the area surveyed due
to the angle of the sheet of galaxies with respect to the plane of the sky
is small (see the formula given in C96),
and that the small scale non-flatness of a ``wall''
(i.e. the velocity dispersion of groups or sparse clusters within
a sheet itself) is also small.  The latter is known to be true in the
Local Universe from analysis
of the CFA redshift survey  
(Dell'Antonio, Geller \& Bothun 1996) as well as of the 
LCRS (Landy \etal\ 1996, Doroshkevich \etal\ 1996).

The difference in co-moving coordinates between adjacent redshift peaks
from the statistically significant sample
in the region of the HDF is given in the penultimate column of
Table 6.   Figure~12 shows
a histogram of these separations, smoothed with a Gaussian with $\sigma = 10$ Mpc.
The smallest separations listed in Table~6 are presumably 
group-to-group separations within a single ``wall'', so the
distribution is displayed both with all the entries listed, and with
the separations less than 50 Mpc omitted.  The vertical lines on
the figure are separated by 68 Mpc. 
While the distribution
of our observed ``wall'' separations shown in Figure~12 is by no means that of
a strictly periodic signal, it does show a characteristic length of
approximately 70 Mpc, with a suggestion that some peaks
have been missed.  This is significantly smaller than is found
through the analysis of local redshift surveys as described above.  
The physical processes that could lead to
such a scale being imprinted on the fluctuation spectrum are
discussed in Szalay (1999).

We caution that pencil beam surveys can exaggerate large scale structure
by emphasizing Fourier components with wave vectors roughly 
perpendicular to the line of sight (e.g. Kaiser \& Peacock 1981).

Between $z=0$ and the maximum redshift at which the redshift peaks
can still be discerned in our survey of the region of the HDF ($z \sim 1.2$),
changing the adopted cosmology among various currently popular
models (ignoring $H_0$, which
just changes the overall scale) distorts the distance scale by
25\%.  If one is sure that the peaks are strictly periodic, an incorrect
choice of cosmological parameters 
would produce blurring of the peaks in Figure~12.
As has already been suggested by
Broadhurst \& Jaffe (1999), this effect could
in principle be used to determine the cosmological parameters,
particularly the cosmological constant,
if one knew that the ``wall'' spacing is strictly periodic and one
knew the period.

A more careful analysis of the implications for large scale
structure of our sample is planned for the future, including
Monte Carlo tests and comparison with pencil beams through
large scale cosmological simulations.

\subsection{Active Galactic Nuclei}

There are three objects with at least one broad emission
line characteristic of QSOs in
this sample.  One is at very high redshift ($z \sim 3.4$).  The other two
are the brightest objects in the peak at $z \sim 0.96$.    These two galaxies,
with $L < 6 L^*$,
are not luminous enough to be considered classical QSOs and are
better designated as broad-lined AGNs.  The single QSO
in the  field studied in Cohen \etal\ (1999a) with $z$ low enough
that the redshift distribution can be determined near $z(QSO)$ is also the brightest
galaxy in a strong redshift peak.  

Our data support the
hypothesis that QSOs/AGNs with $z \gtrsim 0.4$
are in general the brightest
objects in populous groups or clusters of galaxies, as discussed 
for QSOs and radio galaxies with $z < 1$ by Yee \& Green (1984),
see also Yee \& Ellingson (1993).

\section{Luminosity and Density Evolution\label{lumevol}}

We combine our measured redshifts for galaxies in the region
of the HDF with the galaxy models of
Poggianti (1997) to convert the observed
$R$ magnitudes from H99 into luminosities at $R$ in the rest frame.  We use
Poggianti's
passive evolution models with the $z=0$ spectral
energy distributions of present day
ellipticals, Sa and Sc galaxies, assuming them to correspond roughly to
our spectral class ``${\cal A}$'', ``${\cal I}$'' and ``${\cal E}$'' galaxies.
A small extinction correction of $A_R = 0.03$ mag from the maps of
Schlegel, Finkbeiner \& Davis (1998) has been applied. 
We use $L_R^* \equiv -21.75$ mag 
in our adopted cosmology.  This value is 
extrapolated from the results at $B$ compiled by
Binggeli, Sandage \& Tammann (1988).
We do not consider the very
high $z$ regime, where the model galaxy SEDs might be substantially
in error due to strong dependences on the details of the star formation
history adopted. 

Figure~14 displays the luminosity of the galaxies in the region of the HDF in units
of $L_R^*$ as a function of cosmological
co-moving volume rather than of redshift.     
There is no strong increase of $L^*$ with $z$.
A maximum luminosity of 4$L^*$ seems to
cover the full range reasonably well, except for $z < 0.4$, where the volume
sampled is very low and the probability of hitting the brightest galaxies
is correspondingly low, and reduced still further as the HDF was selected
to be devoid of bright galaxies.

Since this is a nearly complete sample, we can examine Figure~12 to
compare the co-moving density of galaxies as a function of $z$.
The completeness limit $R = 23$ is denoted in Figure~12 
for elliptical, Sa and Sc galaxies by a set of curves and it is the area above 
these curves that must be filled in by either the $\sim$8\% of the
$R$-selected sample without redshifts or by 
the EROs that are in this redshift range,
but too faint at optical wavelengths to be included in 
the spectroscopic sample.  A forthcoming paper (Cohen \etal\ 2000) 
will discuss the luminosity function in detail.

\section{The Radio, Mid-IR and Sub-mm Sources in the Region of the HDF\label{radio}}

\subsection{The Radio Sources\label{radiosrc}}

The 8.5 GHz map of Richards \etal\ (1998) in the region
of the HDF is among the
deepest ever made with the VLA, and the positional accuracy for radio sources
in this map is high. 
Bearing in mind the probable astrometric accuracy of the optical coordinates, 
in attempting to find optical counterparts to these radio sources we
impose a positional tolerance of 1.0 arcsec for matching the $R$ catalog from H99
with the VLA secure detections 
with peak flux exceeding 9.0 $\mu$Jy listed in Table~3 of Richards \etal.
If we wish to have a probability of less than 5\% that a match
occurs by chance, the galaxy counts at $R$ imply
that we must restrict the candidates for optical counterparts to
objects with $R < 25.2$.

There are 29 secure VLA sources in the region covered by our
redshift survey.  Using the positional criteria given
above, 15 of these have reasonably bright optical counterparts
with redshifts.  The  difference between the optical
and radio positions
for these objects, which range from $R = 17.7$ to 23.3 mag,
has a mean of 0.1 arcsec in RA and the same in Dec, with
rms dispersions of 0.4 arcsec in each axis, verifying that the
coordinate system of H99 in the Flanking Fields is identical
to that of the VLA.
There is also
one secure identification
of a reasonably bright object ($R = 23.2$ mag)
that does not have a redshift.   There are
two more possible identifications with bright objects
which have positional errors greater than the adopted tolerance
but less than 1.5 arcsec.  

We adopt $4.4 \times 10^{21}$ W Hz$^{-1}$ as the value of the
luminosity of a $L^*$ spiral or irregular
galaxy at 8.5 GHz from the review by Condon (1992).
We calculate $K$-corrections for the 8.4 Ghz flux based on a mean
radio spectral index
of $0.4$ (Richards 2000, see also
Fomalont \etal\ 1991 and Windhorst \etal\ 1993).  We ignore
any change in the radio spectral index with $z$ due to normal stellar evolution.
We then calculate, for the optical counterparts with measured $z$,
the ratio of VLA to optical luminosity, both in units of $L^*$.  
Figure~15 shows this ratio as a function of redshift and it is
also included in Table~7.  
The curve shown in Figure~15 indicates the selection limit imposed by
assuming a galaxy with $L_R = L_R^*$ and with a VLA flux at the minimum
of that for a secure detection, 9 $\mu$Jy.   It produces a reasonably
good definition of the lower envelope of the points representing the
secure optical identifications.  

Details of the identifications
are given in Table~7.  The first column is the optical
ID whose exact coordinates can be found in H99 (or, for objects 
with spectroscopic redshifts, in Table 2), then the redshift, the 
galaxy spectral type and the optical luminosity ($L/L_R^*$).
This is followed by the ratio of radio to optical flux, both approximately in units
of $L^*$ as described above.
The last column gives the difference between the VLA coordinates of the radio
source and the proposed optical counterpart.   If 
the suggested optical counterpart has no redshift, then instead of the luminosity,
the $R$ mag is listed.

For seven of the 29  secure
sources detected by the VLA there is no optical counterpart to $R \sim 25$.
In addition there are four identifications of radio sources
with very faint optical objects with $R \sim 25$ mag,
two of which coincide with ISO sources. 
If the four identifications with very faint objects
are correct and these VLA sources, as well as the ones with
no optical counterpart at all, are real, then these
sources with very faint optical counterparts comprise $\sim$1/3 of the VLA detections at this flux limit.  This is in good agreement with the
independent result of Richards \etal\ (2000) using the Hubble Flanking Field
catalog of Barger \etal\ (1999).

We believe that these extremely faint radio sources represent a 
{\it different} class of object, not a continuation of the
brighter VLA sources towards somewhat fainter optical and radio flux levels.
They cannot simply be more distant or more obscured versions of the sources with
optical counterparts as the gap in properties, particularly in the
$R$ mag of the optical counterparts,
is too large and discontinuous given the small difference in observed radio flux.  

However Barger, Cowie \& Richards (2000) have recently argued that
the rest frame optical IR properties of the radio selected
samples are relatively invariant as a function of redshift.
They suggest that
the faintness of the unidentified galaxies is a consequence
of them lying at higher redshifts where the observed frame 
optical - NIR colors are extremely red and the optical magnitudes faint.

To accomplish this basically requires EROs.  Since the
secure optical counterparts to the VLA radio sources do not 
appear to have significant dust
content,  either there is a discontinuity in their dust content
or these galaxies are passively evolving ``old'' galaxies with $z \sim 1.5$.
In the latter case, the radio emission from these weak sources would
be decoupled from the current star formation rate in the galaxies.

Barger, Cowie \& Richards (2000) discuss the issue of dust at high
redshift in more detail.

\subsection{The Mid-IR Sources}

ISO has surveyed the region of the HDF at 6.5 and at 15$\mu$, with many more
sources detected at the latter wavelength.
The field of ISO is small, 
so observing the region of the HDF requires multiple pointings with multiple spatial
dithers around each pointing.  The observations of
Rowan-Robinson \etal\ (1997) reduced with the algorithms of Aussel \etal\ (1999)
yield 49 sources with a better than 7$\sigma$ detection. 
Aussel \etal\ gave redshifts for 29 of these objects compiled from the
existing literature. All of these lie at $z<1.24$. They also
argued that all of the sources in the HDF proper and all but
six of the sources in the Flanking Fields had secure counterparts and that all
but one of the 8.4 GHz sources from the VLA survey were seen with ISOCAM.

Even with the 7$\sigma$ detection threshold, the positional
uncertainty for ISO sources is worse than with the VLA.  
When we compound that
with the astrometric problems induced by having to tie together multiple
pointings, based on Aussel's evaluation of these uncertainties in the ISO images,
we allow a positional tolerance of 2.5 arcsec for matching
optical and ISO sources.
If we require that the probability that the match occur by chance be less than 5\%, the
candidate optical counterparts must be brighter than $R = 22.8$.

For optical counterparts of ISO sources with 
a spectroscopic redshift, we compute
the emitted luminosity  
from the observed flux seen by ISO with
no $K$-corrections at all and expressed in units of the luminosity of M31 at 12$\mu$,
adopting the IRAS flux of 164 Jy (Rice \etal\ 1988).    We denote this by $L_{M31}(ISO)$.   
The prediction of $K$-corrections for ISO wavelengths
is complicated as several different emission mechanisms contribute
to the mid-IR luminosity of galaxies.  Predictions for starburst galaxies are
given by Elbaz \etal\ (1999); the flux decreases by about a factor of 10 between
$z = 0$ and $z = 1$. 

The results of cross checking the $R$ photometric catalog of H99, 
our redshift catalog and the ISO detections for the region of the HDF
are given in Table~8.  The first column is the optical
ID whose exact coordinates can be found in H99 (or, for objects with spectroscopic redshifts,
in Table 2), then the redshift, the galaxy spectral type and 
the optical luminosity ($L/L_R^*$).
This is followed by the ratio of emitted ISO to optical flux, $L_{M31}(ISO)/[L_R/L_R^*]$.
Then follows the positional difference, whose 
mean in RA is 0.6 arcsec, with 
$\sigma = 1.2$  arcsec, while the mean
difference in Dec is 0.4 arcsec with $\sigma = 1.0$ arcsec.
Figure~16 plots the ratio of $L(ISO)/L(R)$
with the normalization given above as a function of redshift.  The
curve shown in Figure~16 indicates the selection limit imposed by assuming
a galaxy with $L_R = L_R^*$ and with an observed ISO flux at the minimum
of that for a secure detection, 40 $\mu$Jy.  

The large apparent mid-IR luminosities of the ISO sources in the region
of the HDF should not be
a concern.  They are normalized to the mid-IR flux of M31, while 
the IRAS flux at the same wavelength for M82,
the nearest luminous galaxy showing signs of a starburst, given in Rice \etal\ (1988),
is 35 times larger.  

About 2/3 (32/49) of these ISO sources have secure identifications and redshifts.  
We confirm all of the redshifts used by Aussel et al. with the
exception of C36516\_1220 which is at $z=0.401$ rather than $z=0.299$ 
(see the Appendix).  These ISO sources
are matched with the brighter galaxies of the sample at moderate redshift.
Nine more have probable identifications where either the positional tolerance
is within range but the object is fainter than the cutoff of $R = 22.8$ 
or the positional tolerance is slightly higher than the limit.
Two of these have matches with positional differences
of $2.5 < \delta < 3.0$ arcsec for an object with $R < 22.8$  and 
in both of these cases there is a fainter object which is even closer to the ISO position.
Some of the ``probable'' matches may be spurious.

There are only 8 of the 49 ISO objects with no suggested optical counterparts
to $R = 24.5$.  The median flux observed with ISO for the 
unidentified sources is about a factor of 3 smaller than that of the 
identified sources.  A relatively bright optical cutoff ($R = 22.8$ mag)
has been imposed for secure identifications of ISO sources, mandated by the relatively poor
source positions of ISO. Again this is consistent with Aussel \etal\ who
used near IR and optical images and claimed identifications for 43/49
of the sources.

\subsection{The Origin of Emission in the Radio and in the Mid-IR in These Galaxies}

The total dynamic range of both the ISO and VLA data is not large - a
factor of 10 or 20 from
the brightest object detected in their HDF maps to the faintest secure detection.
\footnote{Note that the HDF was selected to exclude bright radio sources.}
Thus there will be a selection effect (i.e. a Malmquist bias)
in looking at any property of
the ISO or radio sources as a function of $z$  as long as this property has
a finite distribution.  In particular, as $z$ increases,
only the most extreme examples will be picked up by the existing radio or mid-IR surveys.
The properties relevant here include luminosity, star formation rate 
(henceforth SFR), 
amount of dust and the
presence of an AGN or of radio emitting gas in a galaxy's environment.  
While a strong AGN will be easily detectable spectroscopically,
a weak AGN may not contribute enough optical luminosity to stand out in 
the total emission from a luminous
galaxy, but will make a noticeable difference in the total radio flux of the galaxy.
Many such examples exist locally, for example M87.

Elbaz \etal\ (1999) review the sources of
emission in the mid-IR, which are thermal emission from dust, 
discrete spectral features emitted (and absorbed) by dust, and 
thermal emission from stars.  Although the vertical axis of Figure~16
is not the SFR, it can be translated into such
by applying a correction scaling factor for the mean SFR per 
unit $R$ luminosity appropriate
for each of the galaxy spectral classes.  Figure~16 does
suggest that the mid-IR emission is monotonically increasing
with SFR, as expected.
Those galaxies whose spectra indicate
relatively little ongoing star formation (i.e. spectra dominated by absorption
lines) appear to emit relatively less flux at the ISO wavelengths
and hence
lie at the lower envelope of the distribution shown in Figure 16.
The AGNs are faint in the mid-IR while the majority of the optical 
counterparts to ISO sources
have galaxy spectral types indicating some or a lot of ongoing star formation.

As reviewed by Condon (1992),
the radio flux for normal galaxies without an active galactic nucleus
arises from free-free emission and from synchrotron emission.  The
radio flux of normal spiral and irregular galaxies is in the mean proportional
to their flux at 60$\mu$.  The models of
Helou \& Bicay (1993) tie the radio flux to the SFR 
through the diffusion of
cosmic-ray electrons whose synchrotron emission produces the radio flux.
However, there is also the possibility
for radio emission from an AGN.  Given that we are selecting the
very brightest sources at each $z$, this cannot be dismissed.

It is clear from comparing Figure~15 with Figure~16
that radio emission is not well correlated with mid-IR emission for these
faint radio sources in the HDF. At least
some of the radio sources must arise from weak AGN, as is
also the case among the optical counterparts 
found by Hammer \etal\ (1995)
for the
microjansky radio sources in one of the CFRS fields. The   
two most radio bright galaxies in the region of the HDF are the
spectroscopically identified AGN C36463\_1404 and the
optically luminous object C36443\_1133, whose optical spectrum 
is that of a classical luminous elliptical galaxy and which may be a cD galaxy
at high redshift.  The latter appears
to show spatially extended radio lobes (Richards \etal\ 1998). 
Most of the galaxies which are detected VLA sources in the
region of the HDF do not show signs of a high current SFR in their spectra.

Machalski \& Condon (1999) have searched the NVSS radio database 
(Condon \etal\ 1998) for galaxies
in the Las Campanas Redshift Survey (Shectman \etal\ 1996) 
of the nearby Universe.  Approximately one third of the detected radio sources
in this flux limited survey 
appear to arise in AGNs rather than in starbursts or normal galaxies. The AGNs 
in the mean had a larger ratio of
radio to optical flux.  It is thus not surprising that
the mid-IR emission seen by ISO
does not correlate well with the radio flux in the distant sources
in the region of the HDF.

To conclude this discussion, we find the fraction of optical counterparts
that are in the statistically complete sample of redshift peaks.
We use the list of peaks and their velocity dispersions in the rest frame
given in Table~6.
For the definite identifications of VLA sources, 11 of the 13
optical counterparts with $z < 1$ lie within 1.5$\sigma$ of a 
peak which is a member of the statistically
significant sample of redshift peaks.  This implies that essentially
all these galaxies lie within the peaks.  The only very
discrepant object (C36414\_1142, at 2.9$\sigma$ from the relevant peak) 
has a very uncertain redshift.
(It is also one of the persistent outliers in the photometric
redshift comparisons.)  For the ISO sources, 26 of the 32
secure optical counterparts have redshifts that are within
1.5$\sigma$ of their respective peak from the statistically complete
sample, implying that 30 of the 32 galaxies lie within the
statistically complete sample of redshift peaks. 

These fractions ($\sim$90\% for the VLA and the ISO sample) 
are even higher than those 
calculated earlier for the field
galaxies (68\%).  The optical counterparts to the VLA and ISO
sources are among the most luminous galaxies in the sample, and so
it is not surprising that they show even more clustering that
the total sample of field galaxies.

\subsection{The SCUBA Sources}

Source identification based
on the poor resolution and near confusion
limited images produced in the sub-mm by SCUBA is a complex problem.  
In this relatively unexplored wavelength regime we also lack the
experience which
might guide us in defining characteristics to choose 
the correct object among the many possible optical candidates within
the typical sub-mm positional error circle. 
The first survey of the HDF in the sub-mm is that of Hughes \etal\ (1998),
who found five sources in a roughly 5 square arcminute region
covering the HDF.  The identification of these objects
has been questioned by 
Richards (1999), who asserts that the SCUBA positions are
systematically shifted on the sky by an amount larger than anticipated by
Hughes \etal\ in their discussion of the accuracy of the SCUBA astrometry.  
Smail \etal\ (1999) suggest that
at least some faint sub-mm galaxies may be EROs,
making a difficult situation even more complex.  
However, a recent analysis by Barger, Cowie \& Richards (2000) of combined
20 cm VLA and SCUBA imaging of the Flanking Fields has shown that
the bulk of the SCUBA sources are very faint in the optical
and near IR ($I>24$ and $K=21-22$) which would place them beyond
the reach of current spectroscopic work.

\section{Summary\label{summary}}

In this paper we have presented our extensive
redshift survey for the HDF and Flanking Fields which contains 671
objects including 610 galaxies and is 92\% complete to $R = 24$ in the HDF
and to $R = 23$  in the Flanking Fields within a diameter of 8 arcmin
centered on the HDF.   
A statistically significant sample of redshift peaks was defined using
Gaussian kernel smoothing combined with a minimum overdensity threshold. 
These
structures have velocity dispersions typical of groups of galaxies.  
In some cases, closely spaced redshift peaks can be decomposed 
into multiple groups of galaxies separated both in
redshift and in position on the sky.
As we have seen before
in C96 and in Cohen \etal\ (1999a), most ($>68$\%)
of the galaxies are located in these peaks,
which are now apparent out to $z \sim 1.2$.
The absorption-line dominated systems preferentially reside in these
redshift peaks and are among the most luminous galaxies in them for $z < 0.8$.   

We interpret the redshift peaks as representing the intersection of the
line of sight of the pencil beam
with ``walls'' of galaxies as originally described in the local universe
from the CFA Survey 
by  de Lapparent, Geller \& Huchra (1986).  Since the characteristic
size of our field is 3 -- 4 Mpc, there is then a fairly large probability that
a group or small cluster of galaxies will lie within our field for most of
these ``walls''.  The spacing of the redshift peaks
along the line of sight corresponds to the characteristic separation of these large
scale structures, which is $\sim$70 Mpc in our adopted cosmology.
The results are consistent in general, although not in detail,
with other analyses of
large redshift surveys in the Local Universe, e.g.
the CFA redshift survey  
(Dell'Antonio, Bothun \& Geller 1996)
and the LCRS (Landy \etal\ 1996, Doroshkevich \etal\ 1996) 
and with those of Broadhurst \etal\ (1990), although these groups find
the ``wall'' separation to be somewhat larger than the value we have found.
At $z \sim 0.1$ this model is supported by the work of the ESO Slice Project
(Vettolani \etal\ 1997) and at
$z \sim 0.3$  by the work of Small \etal\ (1999).
Our survey suggests that these structures originated when the Universe was
less than half its present age.

We have found that about 7\% of the $K$-selected sample ($K < 20$) are
extremely red objects with $R - K > 5$ and the reddest such
has $R - K = 6$.  One of these has a redshift
and is a galaxy at $z \sim 1$ with no detectable current star formation.
We suspect that this is typical of all the EROs in this magnitude 
and color range.

We have searched for optical counterparts of the sources detected
with the VLA at 8.5 GHz by Richards \etal\ (1998).  Two thirds of
the secure radio sources with observed fluxes at this wavelength exceeding
9 $\mu$Jy can be identified with bright galaxies ($R \sim 21$) 
at moderate redshift.  About 25\% of them have no optical counterpart
or a very faint one ($R \ge \sim25$).  These appear to be a 
different population from the first group rather than just
a continuation of them in some property such as distance or dust content.

We have also searched for optical counterparts to 
the ISO sources of Aussel \etal\ (1999).  About 2/3 of these can be identified
with  bright galaxies ($R \sim 21$ mag) at moderate redshift.  Because
of the larger positional uncertainty of the ISO astrometry and the
high areal density of faint optical objects, the nature of
the ISO sources without bright optical counterparts 
cannot be determined at this time, but is consistent with them
mostly being galaxies somewhat fainter than the adopted brightness cutoff.

The presence of emission in the mid-IR appears to be roughly proportional 
to the rate of ongoing star formation, as would be expected. 
However, the presence of detectable radio emission is
not so tightly coupled to the current star formation rate; a substantial
fraction of these radio sources must be weak AGNs.  

Both the VLA and the ISO sources appear to be even more clustered than
the sample as a whole, with $\sim$90\% of them lying
within the statistically complete sample of redshift peaks.

We have also used our catalog of redshifts in the region of the HDF to demonstrate
that photometric redshift schemes can predict redshifts to high precision
(as good as $\sigma = 5\%$ in $(1+z)$) for the majority of galaxies 
with $z < 1.3$.
 
\appendix
\section{Changes to Published Redshifts}

The sample of the Lick Deep Group is used as published in Phillips \etal\ (1997)
and Lowenthal \etal\ (1997) with the addition of one galaxy whose
spectrum was obtained in 1996 but no redshift was deduced until early 1999
(H36384\_1231).  Also data from the setup stars (all of which
are relatively bright stars) used to align the slitmasks
were incorporated into the redshift table.
As communicated to JGC by A. Phillips,
the redshift of F36380\_0922 (published as $z = 0.512$ in
Phillips \etal\ 1997) was changed to $z = 0.767$ .

The redshifts given by Steidel \etal\ (1996) 
were used as modified and supplemented
in the review by Dickinson (1998).  Note that one redshift was changed
in the latter (H36482\_1417, published originally as 2.845
and corrected later to 2.008) and one (H36513\_1227)
was withdrawn completely.

Several modifications were made to the redshifts published in C96.
In most cases, these were not because the features original seen were not really
present, but rather because the original interpretation of the features in the spectrum
was subsequently shown to be incorrect.  
The redshift of F36528\_1453 was changed when it was realized that the
line previously identified as 3727 \AA\ of [OII] is actually the 5007 \AA\ line
of [OIII].
The redshift of F36559\_1454, which was initially
thought to be a late type galactic star, changed when it was realized
that the feature originally ascribed to MgH was actually the H+K break.
The redshift of F37098\_1523 changed when spectra reaching further towards
the blue became available which clearly showed the H+K break
to be bluer than the spectral coverage of the earlier spectra.
The new spectrum indicated that the feature ascribed
to the H+K break was actually the Mg triplet region.
The redshift of H36446\_1227 changed when it was realized that the far-UV
interstellar lines had been misidentified initially.
The original redshift of H36516\_1220 is from
a spectrum of low SNR.  A better spectrum subsequently obtained by
the Hawaii group led to a redetermination of the redshift of this
object.  In addition to these substantive changes, a small 
modification was made to the redshift of F36223\_1241.
Table~9 lists the six corrections to redshifts published in C96
made here.  

In addition, there was a case of object confusion in C96 
among the pair
H36467\_1236 and H36470\_1236 which has been corrected here.

Updates to the redshifts originally from the Hawaii web site 
are not listed here.
Similarly, some of the redshifts 
published as preliminary values in Hogg \etal\ (1998)
have been modified; see the notes in the redshift catalog
of Table 2.

\section{The Objects With $R < 24$ in the HDF Without Redshifts}

Table~10 lists the eight objects that are present in the
photometric survey of H99 with $R < 24$, are in the HDF itself, and
do not have spectroscopic redshifts at this time. 
Most of these objects have been observed spectroscopically
with the LRIS at Keck Observatory more than once each, and
for several hours each time.

NOTE ADDED MAY 17, 2000 -- There are two new redshifts for objects
in the HDF with $R < 24$, H36377\_1235 and H36472\_1342.  The former
comes from my continuing observations in this region, while the
latter is from S.Dawson, D.Stern, A.J.Bunker, H.Spinrad, \& A.Dey (2000,
in preparation).  There are also two new redshifts for objects
with $z \sim 4.5$ from Stern \& Spinrad (1999).
These four objects are listed
in Table 2b, but are not included anywhere else in this paper.  A note
in press will be added to the published version as well with these
new redshifts, although it is probably too late to incorporate them
into Table 2b in the paper which will appear shortly in the ApJ.

\acknowledgements The entire Keck/LRIS user community owes a huge debt
to Jerry Nelson, Gerry Smith, Bev Oke, and many other people who have
worked to make the Keck Telescope and LRIS a reality.  We are grateful
to the W. M. Keck Foundation, and particularly its late president,
Howard Keck, for the vision to fund the construction of the W. M. Keck
Observatory.  

We are very grateful to A. Phillips and other members of the 
Lick Deep Group for permission
to examine their spectra of objects in the region of the HDF and
for their cooperation in resolving problem cases.
We are also very grateful to Brad Behr for help in constructing Figures~2 and 3.

JGC is grateful for the hospitality of Princeton University and RBD and JGC
are grateful for the hospitality of the Institute for Advanced Study.

JGC is grateful for partial support from STScI/NASA grant AR-06337.12-94A.
RDB acknowledges support under NSF grant AST95-29170.
DWH was supported in part by a Hubble Fellowship grant
HF-01093.01-97A from STScI (which is operated 
by AURA under NASA contract NAS5-26555).
Kevin Richberg is grateful to the Caltech SURF program for partial support.

\clearpage 
%
%
\begin{deluxetable}{lrr}
\tablenum{1}
\tablewidth{0pt}
\tablecaption{Sources for the Redshifts}
\label{tab1}
\tablehead{\colhead{Reference} & \colhead{Number Objects} & 
\colhead{Abbrev.\tablenotemark{a}} \nl
}
\startdata
Previously Published: \nl
Cohen \etal\ (1996)  (Caltech + Hawaii)   &           140  &   c96  \nl 
Phillips \etal\ (199)+Lowenthal \etal\ (1997)   &    129  &  l96 \nl
Steidel \etal\ (1996) (See also Dickinson 1998)  &   15 &   s96 \nl
Steidel \etal\ (1999)                 &         2  &   s99 \nl
Spinrad \etal\ (1998)     &              1   &  spi \nl
Waddington \etal\ (1999) & 1 & wad \nl
Weymann \etal\ (1998)      &       1   & wey \nl
~ \nl
This Paper: \nl
Cohen         &             $\sim$380 & cal \nl
Hawaii        &      $\sim$150 & haw \nl
Steidel       &   5  &   st0 \nl 
\enddata
\tablenotetext{a}{Abbreviation used in Table 2}
\end{deluxetable}

%
%
\begin{deluxetable}{rrrrrrrclr}
\tablenum{3}
\tablewidth{0pt}
\tablecaption{Objects Slightly Outside the Sample Area}
\label{tab3}
\tablehead{\colhead{RA} & \colhead{} & \colhead{Dec\tablenotemark{a}} &
\colhead{} & 
\colhead{$R$} & \colhead{$z$} & \colhead{Quality} &
\colhead{Sp.Type} & \colhead{Source\tablenotemark{b}} \nl
\colhead{ ($-12$h)} & \colhead{} &
\colhead{($-62^{\circ}$)} & \colhead{} &
\colhead{(mag)} \nl
}
\startdata
37 &  3.29  &  16 & 46.6 &  22.27 & 0.744  &   1 & ${\cal EI}$ & c96 \nl
37 &  4.61  &  16 & 51.8 &  20.15 & 0.377  &   1 & ${\cal I}$ &  c96 \nl
37 &  8.12  &  16 & 31.4 &  21.94 & 0.000  &   1 & ${\cal M}$  & c96 \nl
37 &  8.15  &  16 & 59.5 &  20.54 & 0.458  &   1 & ${\cal I}$ &  c96 \nl
37 & 10.71  &   9 & 34.8 &  24.62 &  4.129 &   ? & $\cal X $ & s99  \nl
37 & 11.99  &  16 & 59.6 &  22.86 & 1.142  &   1 & ${\cal EI}$ & c96 \nl
37 & 14.38  &  15 & 59.3 &  21.22 & 0.567  &   4 & ${\cal E}$ &  cal \nl
37 & 16.49  &  16 & 30.0 &  22.51 & 0.000  &   1 & ${\cal M}$ &  c96 \nl
37 & 16.67  &  16 & 44.4 &  21.98 & 0.557  &   1 & ${\cal EI}$ & c96 \nl
37 & 16.82  &  10 &  7.4 &  20.29 & 0.411  &   1 & ${\cal I}$ &  haw \nl
37 & 18.45  &  15 & 53.7 &  20.65 & 0.476  &   1 & ${\cal I}$ &  c96 \nl
37 & 18.78  &  16 &  4.8 &  21.72 & 0.558  &   1 & ${\cal EI}$ & c96 \nl
37 & 20.62  &  11 &  5.7 &  24.11 &  4.421 &   2 & $\cal A $ & s99  \nl
37 & 22.24  &  11 & 24.1 &  22.33 & 0.786  &   1 & ${\cal E}$ &  l96,cal \nl
37 & 22.45  &  16 & 12.9 &  21.87 & 0.000  &   1 & ${\cal M}$ &  c96 \nl
\enddata
\tablenotetext{a}{J2000 coordinates}
\tablenotetext{b}{The abbreviations used to denote redshift sources
are listed in Table 1.}
\end{deluxetable}

%
%
\begin{deluxetable}{crrrrrrr}
\tablenum{4}
\tablewidth{0pt}
\tablecaption{Median $z$ in Bins of $R$ and $K$ Mag}
\label{tab4}
\tablehead{\colhead{Mag Bin} & \colhead{No.} & \colhead{Median} &
\colhead{Q1} & \colhead{Q3} & \colhead{Median\tablenotemark{a}} & 
\colhead{Q1\tablenotemark{a}} & \colhead{Q3\tablenotemark{a}} \nl
}
\startdata
$R$ \nl
 16.5 -- 17.0 & 1 & 0.11 & ... & ... \nl
 17.0 -- 17.5   &   1  & 0.14 &  ... & ... \nl
 17.5 -- 18.0 &   3 &  0.08 &  ... & ... \nl
 18.0 -- 18.5 &  0 \nl
 18.5 -- 19.0   &   5  &   0.14    & ... & ... \nl
 19.0 -- 19.5   &   5 &   0.20   & ... & ... \nl
 19.5 -- 20.0   &   8 & 0.42 & ... & ... \nl
 20.0 -- 20.5  &   24  &   0.47 &  0.52 &  0.38 \nl
 20.5 -- 21.0  &   47   &  0.46 &       0.52 &  0.32 & \nl
 21.0 -- 21.5  &   54  &   0.51 &       0.56 &  0.47 \nl
 21.5 -- 22.0  &   76  &   0.57 &       0.78 &  0.48 \nl
 22.0 -- 22.5  &  94  &  0.71 &   0.89 &  0.53  \nl
 22.5 -- 23.0 &  112   &  0.68 &       0.89 &  0.48 & 0.74 &   0.90 &  0.48 \nl
 23.0 -- 23.5  &   70  &    0.82 &    1.01 & 0.51 &  0.88 & 1.19 & 0.52 \nl
 23.5 -- 24.0  &  49  &  0.79  &     1.09 &  0.48 & 1.02 & 1.60 & 0.55 \nl
~ \nl
$K$ \nl
  14.5 -- 15.0  &    2  &   0.11  \nl
  15.0 -- 15.5   &   2  &   0.11 \nl      
  15.5 -- 16.0   &   3  &   0.08  \nl
  16.0 -- 16.5  &    5  &   0.23 \nl
  16.5 -- 17.0  &   18  &   0.52  \nl
  17.0 -- 17.5  &   28   &  0.50 \nl
  17.5 -- 18.0  &   42  &   0.56 \nl
  18.0 -- 18.5  &   63  &   0.56  \nl
  18.5 -- 19.0  &   56   &  0.58 \nl
  19.0 -- 19.5  &   65   &  0.56 \nl
  19.5 -- 20.0   &  62  &   0.68 \nl
  20.0 -- 20.5   &  64  &    0.68 \nl
  20.5 -- 21.0  &   23  &   0.68 \nl 
\enddata
\tablenotetext{a}{Corrected for $z \sim1.7$ galaxies missing from sample, see text.}
\end{deluxetable}

%
%
\begin{deluxetable}{lrrlrrrl}
\tablenum{5}
\tablewidth{0pt}
\tablecaption{Tests of Photometric Redshifts}
\label{tab5}
\tablehead{\colhead{~} & \colhead{$z < 1.5$}  & \colhead{~} &
\colhead{~} & \colhead{~} & \colhead{$z > 1.9$} &  \colhead{~}  \nl
\colhead{No.} & \colhead{Mean$\Delta$\tablenotemark{a}} & 
\colhead{$\sigma$\tablenotemark{a}} & \colhead{Corr.Coef} &
\colhead{No.} & \colhead{Mean$\Delta$\tablenotemark{a}} & 
\colhead{$\sigma$\tablenotemark{b}} & \colhead{Corr.Coef} \nl
}
\startdata
Lanzetta \etal\ \nl
33 (all) & 0.15 & 0.30  & 0.47 & 8 & $-$0.20 & 0.22 & 0.22 \nl
28 ($-$outliers) & 0.05 & 0.82 & 0.10 & 7 & $-$0.13 & 0.09 & 0.89 \nl
~ \nl
Sawicki \etal\ \nl
40 (all) & 0.04 & 0.21 & 0.46 & 9 & 0.01 & 0.10 & 0.90 \nl
36 ($-$outliers) & $-$0.01 & 0.06 & 0.92 \nl
~\nl
Wang \etal\ \nl
24 (all) & 0.08 & 0.26 & 0.39 & 12 & $-$0.03 & 0.32 & 0.09 \nl
20 ($-$outliers) & $-$0.02 & 0.04 & 0.95 & 10 & 0.09 & 0.14 & 0.78 \nl
\enddata
\tablenotetext{a}{Mean and $\sigma$ of $[(z_{sp} - z_{phot})/(1+z_{sp})]$.}
\end{deluxetable}

%
%
\begin{deluxetable}{lrrrr}
\tablenum{6}
\tablewidth{0pt}
\tablecaption{Velocity Dispersions for the Strong Redshift 
Peaks in the Region of the HDF}
\label{tab6}
\tablehead{\colhead{$z(peak)$\tablenotemark{a}} & \colhead{N(peak)\tablenotemark{b}} & 
\colhead{Co-Moving Distance} & \colhead{$\Delta(D)$} &
\colhead{$\sigma(v)(N)$} \nl 
\colhead{} & \colhead{} & \colhead{(Mpc)\tablenotemark{c}} & 
\colhead{(Mpc)\tablenotemark{c}} & 
\colhead{(km s$^{-1}$)\tablenotemark{d}} \nl
}
\startdata
0.079\tablenotemark{e} & 3 & 380 & ... & 260 \nl   
0.086\tablenotemark{e} & 5 & 410 &  30 & $\le$150 \nl
0.107 & 5 & 555 & 145 & 750 \nl
0.137 & 7 & 635 &  80 & 375 \nl
0.201 & 4 & 905 & 270 & 255 \nl
0.253 & 4 & 1115 & 210 & 305 \nl
0.321 & 14 & 1370 & 255 & $\le$150 \nl
0.409\tablenotemark{f} & 13 &  1690 & 320 & $\le$150 \nl
0.421\tablenotemark{f} & 11 & 1730 & 40 & 335 \nl
0.457 & 19 & 1850 & 120 & 315 \nl
0.475\tablenotemark{e} & 35 & 1915 & 65 & 325 \nl
0.485\tablenotemark{e} & 19 & 1945 & 30 & 355 \nl
0.516 & 36 & 2045 & 100 & 705 \nl
0.560 & 45 & 2190 & 145 & 640 \nl
0.681 & 20 & 2555 & 365 & 565 \nl
0.749 & 32 & 2755 & 200 & 745 \nl
0.848 & 35 & 3030 & 275 & 595 \nl
0.936 & 14 & 3260 & 230 & 445 \nl
0.960 & 16 & 3325 & 65 & 615 \nl
1.016 & 26 & 3465 & 140 & 360 \nl
1.141 & 4 & 3770 & 305 & 185 \nl
1.221 & 5 & 3960 & 190 & 390 \nl
1.243 & 4 & 4010 & 50 & 370 \nl
\enddata
\tablenotetext{a}{Statistically significant sample of peaks as defined in text.}
\tablenotetext{b}{Number of members calculated assuming a Gaussian distribution.}
\tablenotetext{c}{Adopted cosmology is $H_0 = 60$~\kmsM and 
$\Omega_M = 0.3$ with $\Lambda = 0$.}
\tablenotetext{d}{$\sigma$ is given in the rest frame.
The instrumental error of 200 km s$^{-1}$ in the observed
frame has been removed in quadrature in the rest frame.}
\tablenotetext{e}{These close peaks were resolved by the group 
finding algorithm.}
\tablenotetext{f}{This pair of close peaks were split by hand.}
\end{deluxetable}


%
%
\begin{deluxetable}{lrr}
\tablenum{9}
\tablewidth{0pt}
\tablecaption{Modifications to Redshifts Published in Cohen \etal\ (1996)}
\label{tab9}
\tablehead{\colhead{Galaxy} & \colhead{$z$ (c96)} & 
\colhead{$z$ (This paper)} \nl
}
\startdata
F36223\_1241 &  0.498 & 0.458 \nl
F36528\_1453 &  0.463 & 0.089 \nl
F36559\_1454 &  0.000 & 0.321 \nl
F37098\_1523 &  0.597 & 0.200 \nl
H36446\_1227 &  2.268 & 2.500 \nl
H36516\_1220 &  0.299 & 0.401 \nl
\enddata
\end{deluxetable}

%
%
\begin{deluxetable}{lrlr}
\tablenum{9}
\tablewidth{0pt}
\tablecaption{HDF Objects With $R < 24$ and With No Redshift}
\label{tab10}
\tablehead{\colhead{Object} & \colhead{$R$} & 
\colhead{Object} &\colhead{$R$} \nl
\colhead{} & \colhead{(mag)} & \colhead{} & \colhead{(mag)} \nl
}
\startdata
H36378\_1235 & 23.9 & H36398\_1229 & 23.2 \nl
H36453\_1143 & 24.0 & H36472\_1342 & 23.9 \nl
H36485\_1317 & 23.4 & H36526\_1202 & 23.4 \nl
H36530\_1344 & 24.0 & H36533\_1222 & 23.7 \nl
\enddata
\end{deluxetable}

\clearpage

\begin{figure}
\epsscale{0.7}
\plottwo{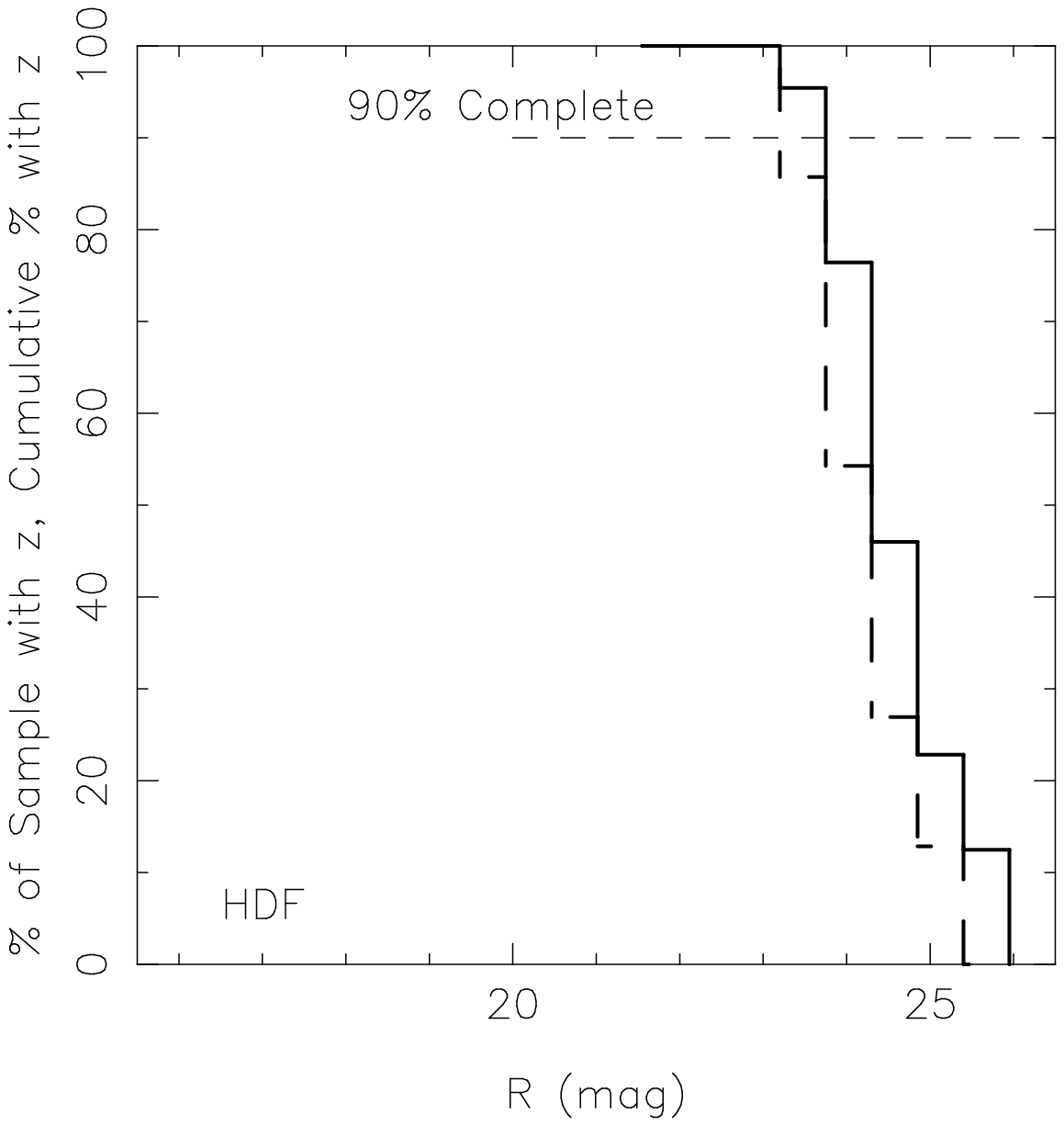}{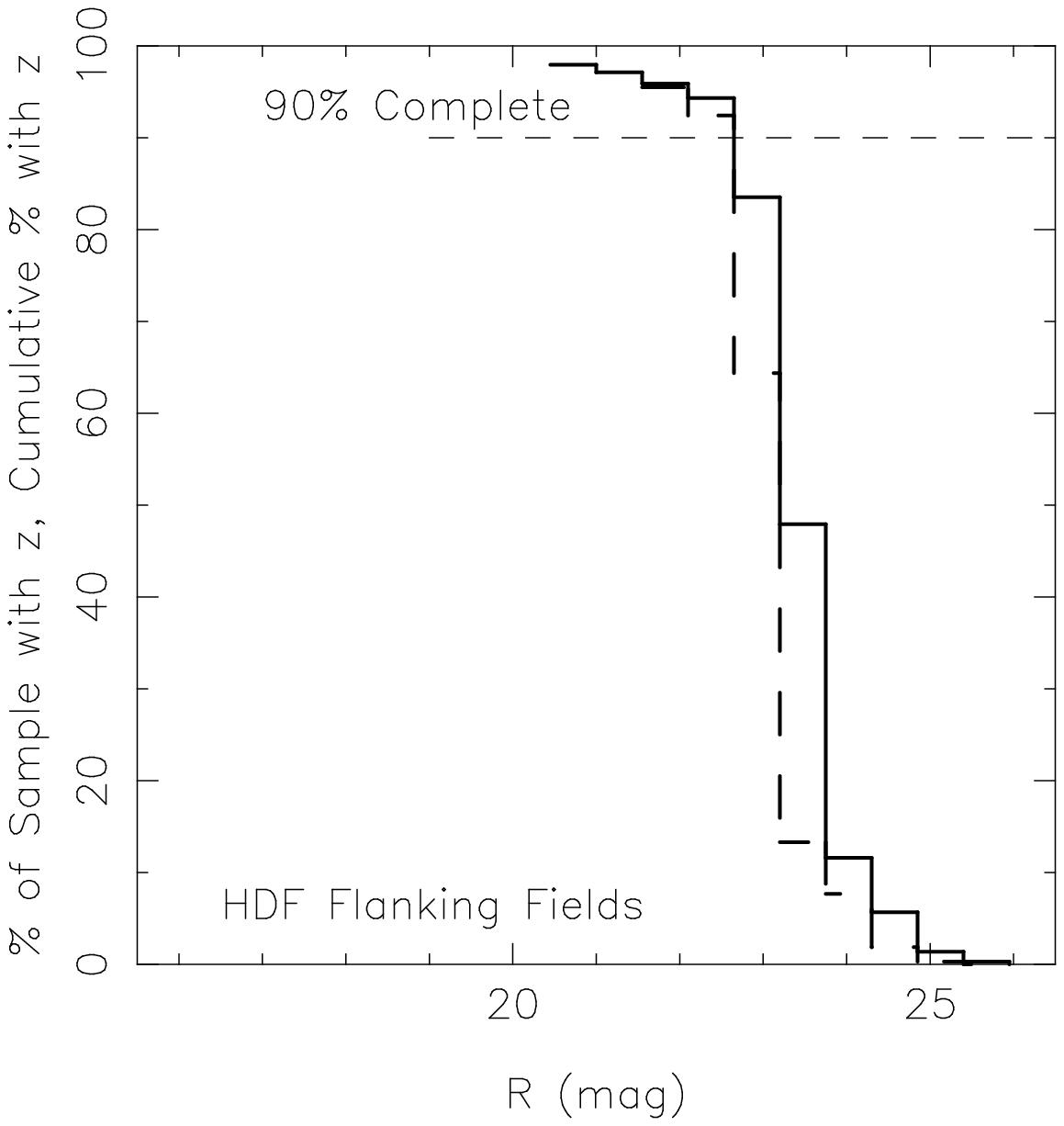}
\caption[figure1.ps]{The redshift completeness in the HDF (Figure 1a) and in
the Flanking Fields (Figure 1b) is shown as a function of $R$ mag.  The dashed
line represents the completeness in 0.5 mag bins, while the solid line represents
the cumulative completeness.  The horizontal line indicating 90\% 
completeness is also indicated.
\label{fig1}}
\end{figure}

\clearpage

\begin{figure}
\epsscale{0.7}
\caption[figure2.ps]{The redshift survey objects are indicated, together
with their redshifts, superposed on an image of the HDF
taken, with permission, from Williams \etal\ (1996).  For $z \le 1$, galaxies
of spectral class ``${\cal E}$'' are indicated by blue circles, while
green circles denote ``${\cal I}$'' galaxies, and red circles denote
absorption line ``${\cal A}$'' galaxies.  White circles are used for
objects with $z \ge 1$ and for 
``${\cal X}$'' galaxies.
\label{fig2}}
\end{figure}

The Postscript file for figure 2 is very large. It is available
via anonymous ftp 
from phobos.caltech.edu/$\sim$ftp/users/jlc/hdf\_image/hdf\_redshift.ps.gz.
Once this article is published in the electronic edition of the ApJ,
Figure 2 will have to be obtained from that source, and this file will
be removed from phobos.

\clearpage

\begin{figure}
\epsscale{1.4}
\caption[figure3.ps]{(a of a,b,c,d) The redshift survey objects are indicated, together
with their redshifts, superposed on an image of the Flanking Fields around the
HDF, broken into four quadrants.  The HST image is
taken, with permission, from Williams \etal\ (1996).  The same color coding is
used to indicate the spectral classes of galaxies with $z \le 1$ as in 
Figure~2,
\label{fig3}}
\end{figure}

The Postscript file for figure 3a,b,c,d is too large for the Astrooph server.  It is available
via anonymous ftp 
from phobos.caltech.edu/$\sim$ftp/users/jlc/hdf\_image.
Once this article is published in the electronic edition of the ApJ,
Figure 2 will have to be obtained from that source, and this file will
be removed from phobos.

\clearpage

\setcounter{figure}{2}
\begin{figure}
\epsscale{1.4}
\caption[figure3.ps]{(b of abcd) The redshift survey objects are indicated, together
with their redshifts, superposed on an image of the Flanking Fields around the
HDF, broken into four quadrants.  The HST image is
taken, with permission, from Williams \etal\ (1996).  The same color coding is
used to indicate the spectral classes of galaxies with $z \le 1$ as in 
Figure~2,
\label{fig3}}
\end{figure}

\clearpage

\setcounter{figure}{2}
\begin{figure}
\epsscale{1.4}
\caption[figure3.ps]{(c of abcd) The redshift survey objects are indicated, together
with their redshifts, superposed on an image of the Flanking Fields around the
HDF, broken into four quadrants.  The HST image is
taken, with permission, from Williams \etal\ (1996).  The same color coding is
used to indicate the spectral classes of galaxies with $z \le 1$ as in 
Figure~2,
\label{fig3}}
\end{figure}

\clearpage

\setcounter{figure}{2}
\begin{figure}
\epsscale{1.4}
\caption[figure3.ps]{(d of abcd) The redshift survey objects are indicated, together
with their redshifts, superposed on an image of the Flanking Fields around the
HDF, broken into four quadrants.  The HST image is
taken, with permission, from Williams \etal\ (1996).  The same color coding is
used to indicate the spectral classes of galaxies with $z \le 1$ as in 
Figure~2,
\label{fig3}}
\end{figure}

\clearpage

\begin{figure}
\epsscale{0.8}
\plotone{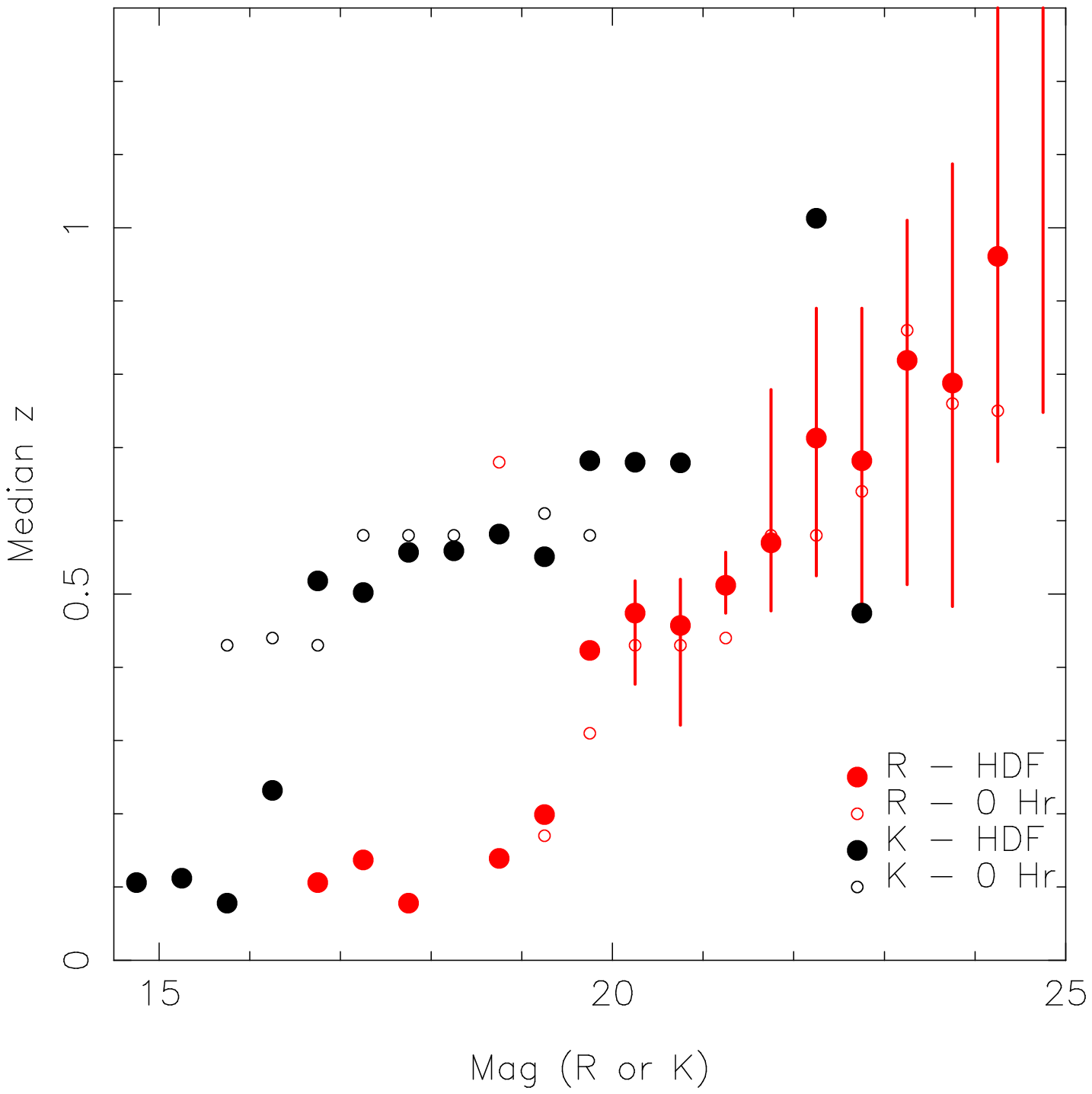}
\caption[figure4.ps]{The median redshift is shown as a function of the $R$
and of the $K$ mag for the sample in the region of the HDF and for the sample
in the field J0053+1234 from Cohen \etal\ (1999a).  
The first and last quartiles for the HDF sample are indicated
for $R$ by error bars only when there are more than 10 galaxies in a bin.
\label{fig4}}
\end{figure}

\clearpage

\begin{figure}
\epsscale{0.8}
\plottwo{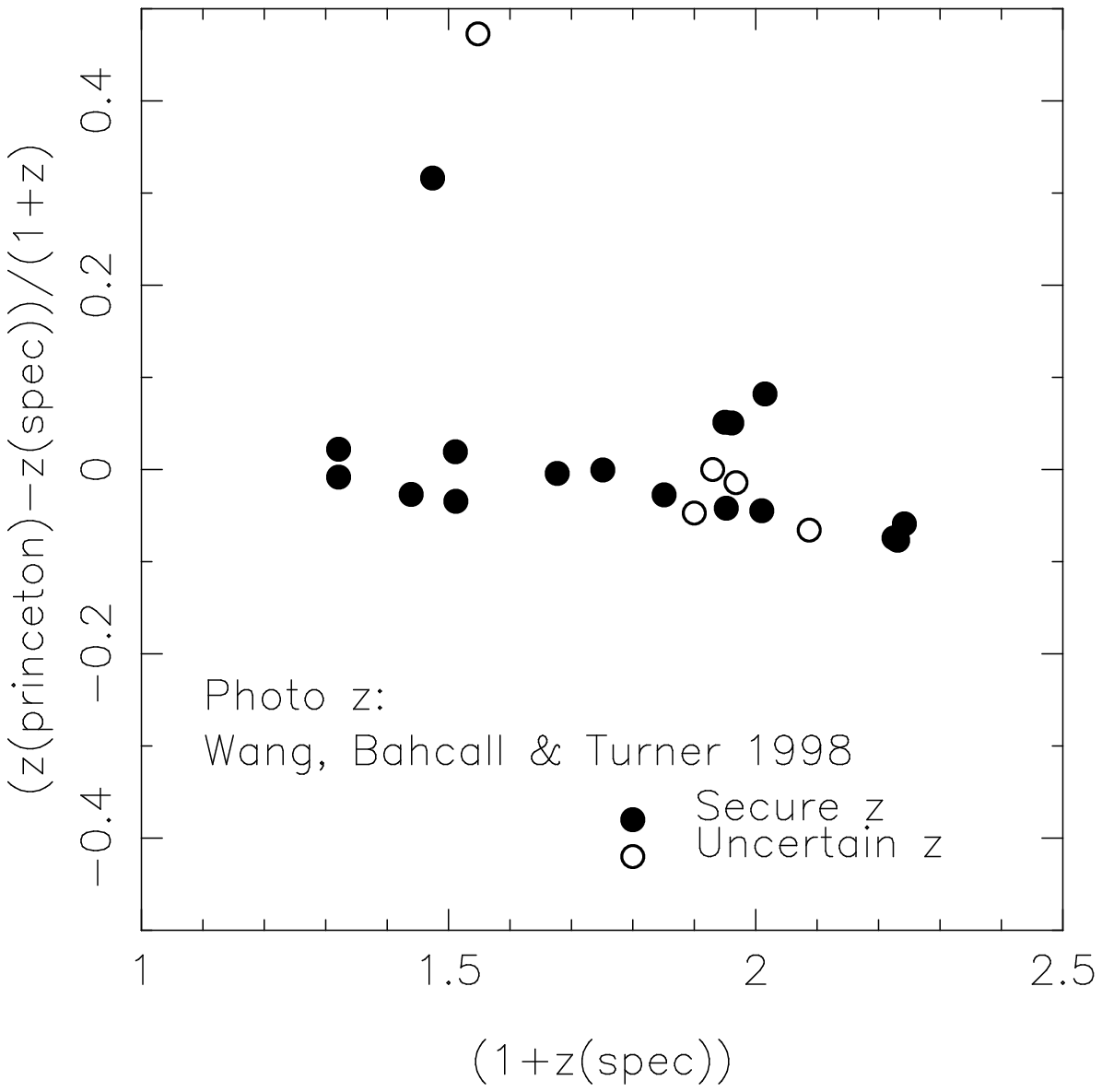}{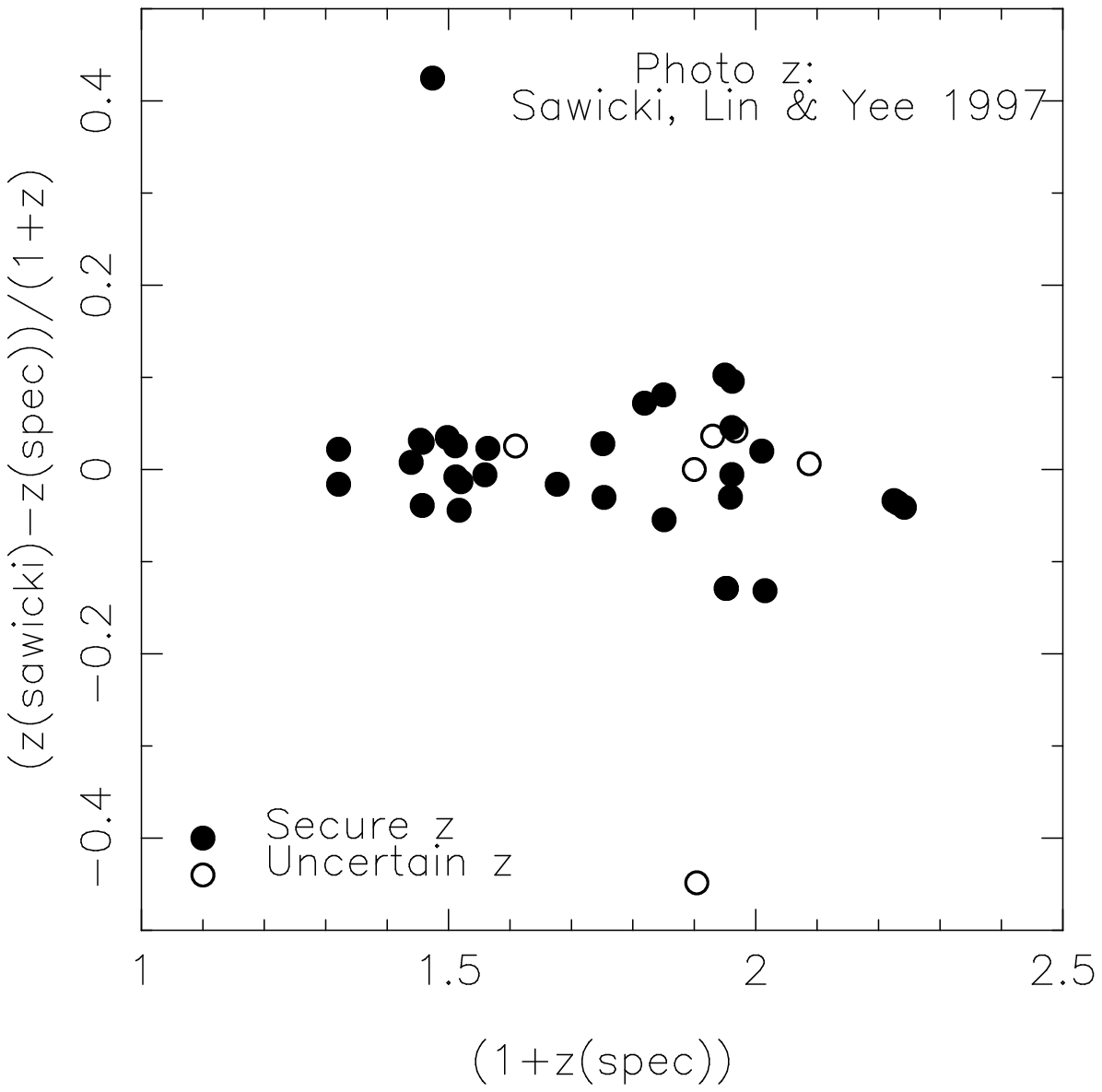}
\epsscale{0.6}
\plotone{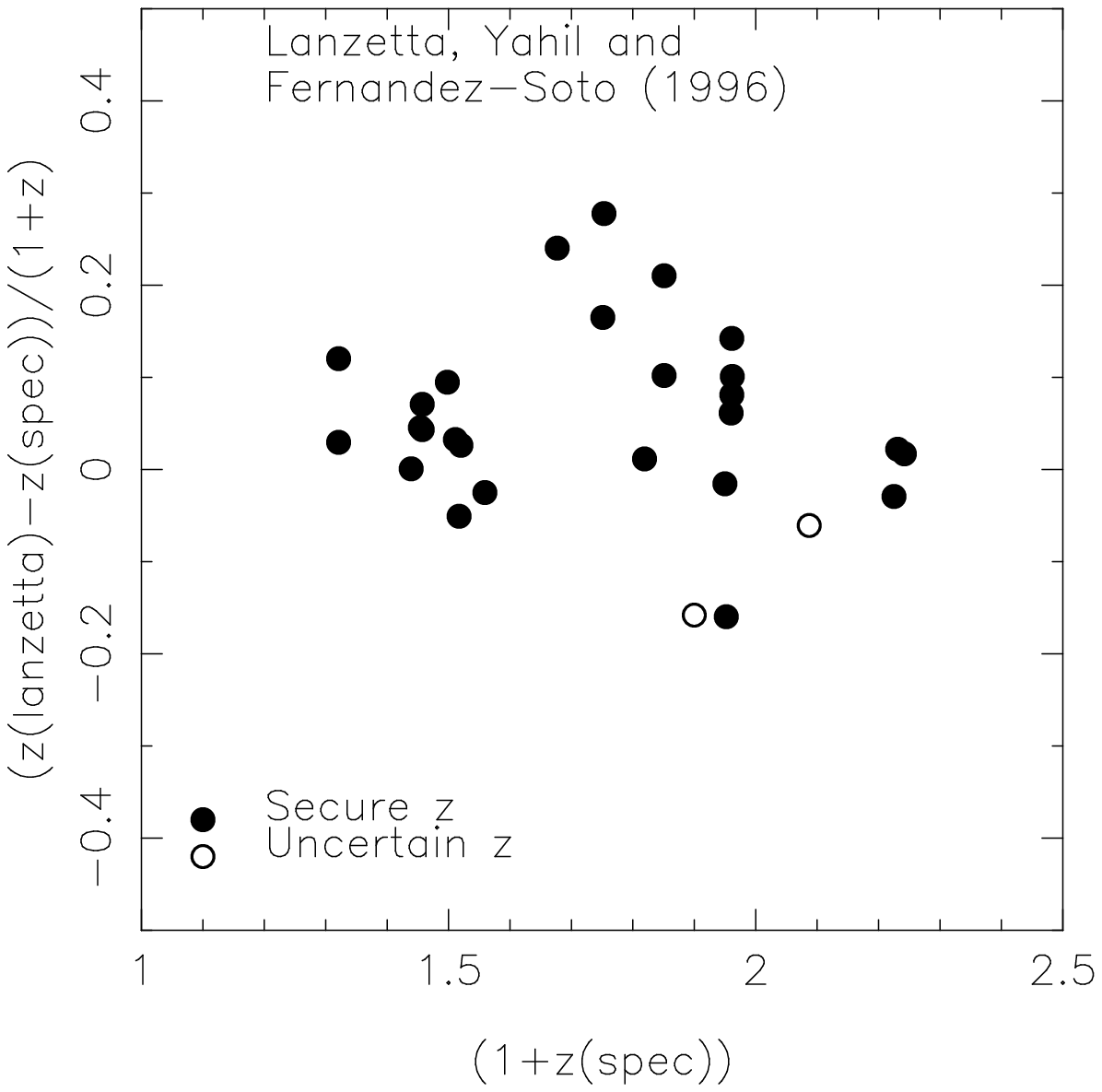}
\caption[figure5.ps]{The predictions of the three photometric
redshift techniques are tested against the observed spectroscopic
redshifts in the region of the HDF.  Filled circles denote secure
spectroscopic redshifts, while open circles indicate less certain
spectroscopic redshifts.
\label{fig5}}
\end{figure}

\clearpage

\begin{figure}
\epsscale{0.7}
\plotone{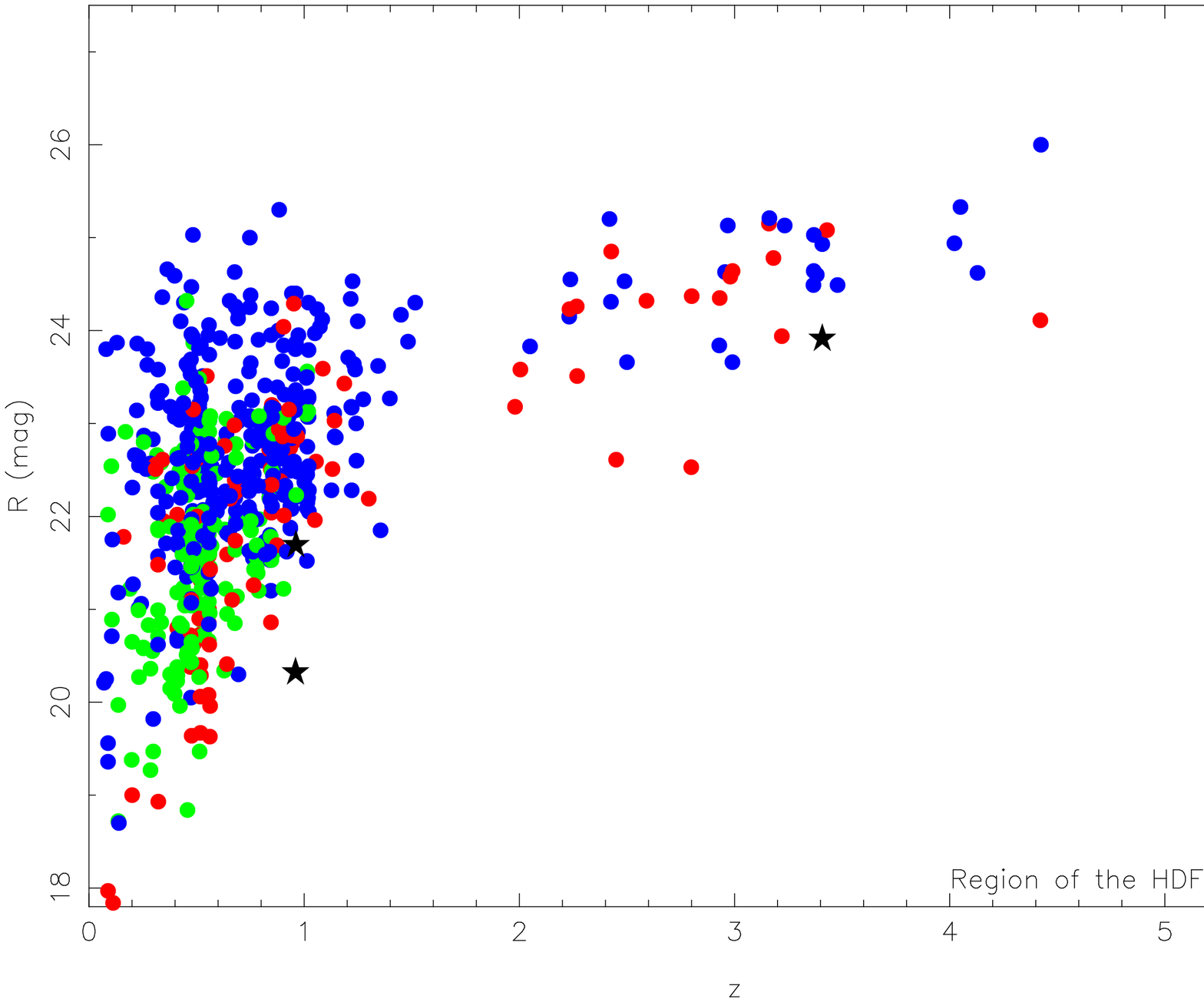}
\caption[figure6.ps]{The $R$ mag is shown as a function of redshift
for galaxies in the region of the HDF.  Blue circles denote ``${\cal E}$'' galaxies,
green circles denote ``${\cal I}$'' galaxies, while red circles denote
absorption line ``${\cal A}$'' galaxies.  QSOs/broad lined AGNs
are indicated by black stars.
\label{fig6}}
\end{figure}

\begin{figure}
\epsscale{0.8}
\plotone{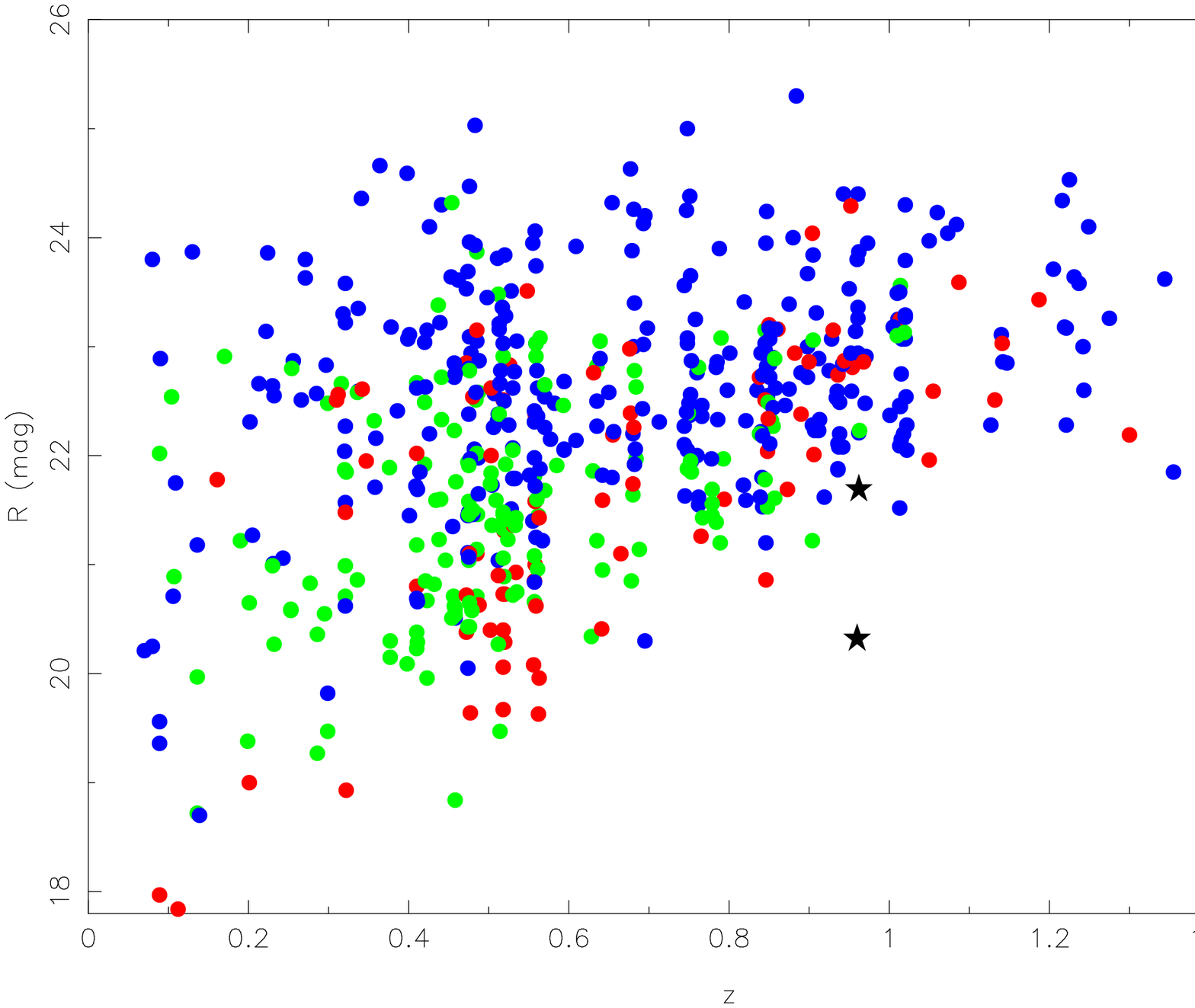}
\caption[figure7.ps]{The $R$ mag is shown as a function of redshift
for galaxies in the region of the HDF for $z < 1.6$.  The
symbols are the same as those of Figure~6.
\label{fig7}}
\end{figure}

\begin{figure}
\epsscale{0.7}
\plotone{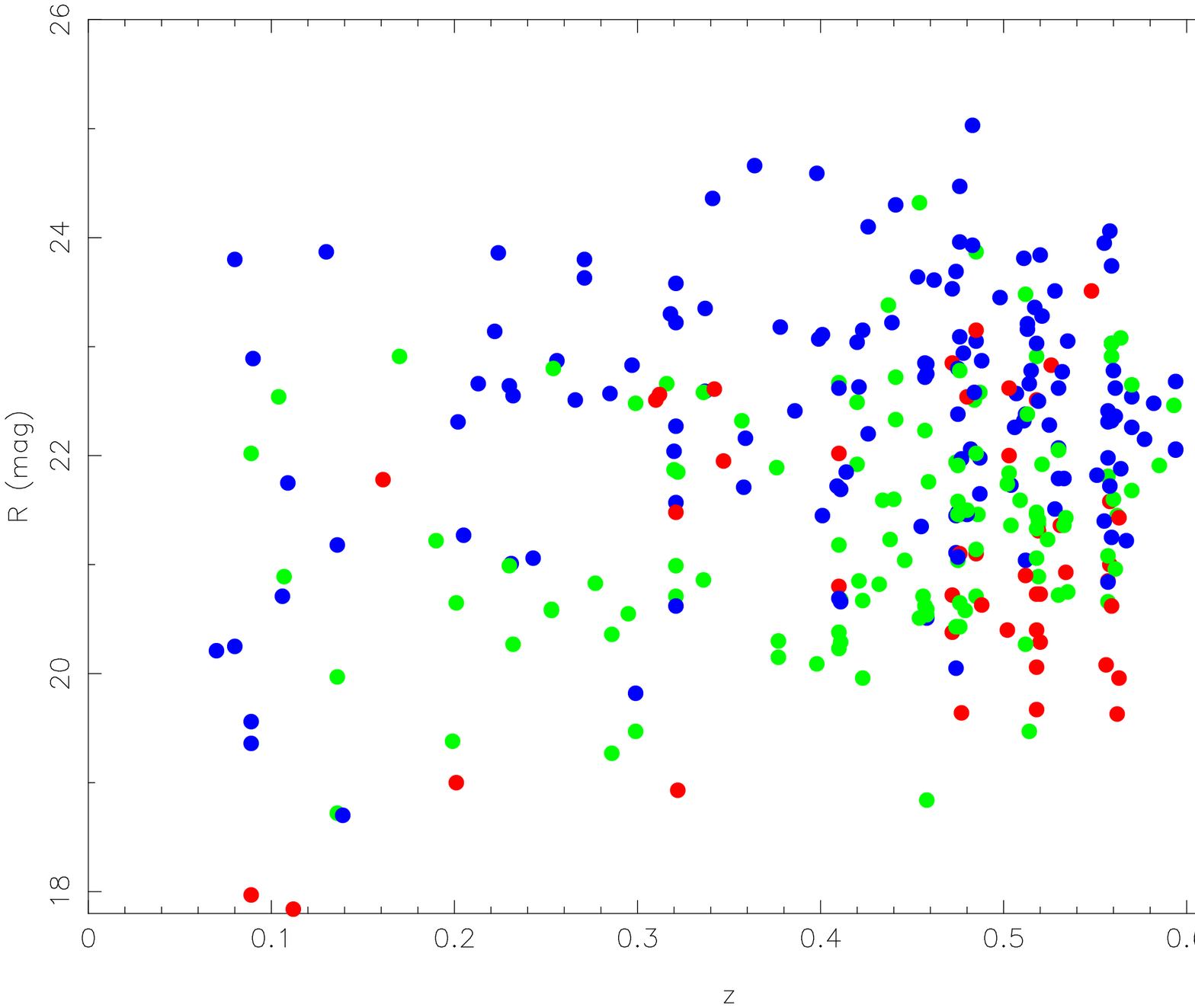}
\caption[figure8.ps]{a) The $R$ mag is shown as a function of redshift
for galaxies in the region of the HDF for $z < 0.7$ (figure 8a) and for
$0.6 < z < 1.4$ (figure 8b).  The symbols are the same as those of Figure~6.
\label{fig8}}
\end{figure}

\clearpage

\setcounter{figure}{7}
\begin{figure}
\epsscale{0.7}
\plotone{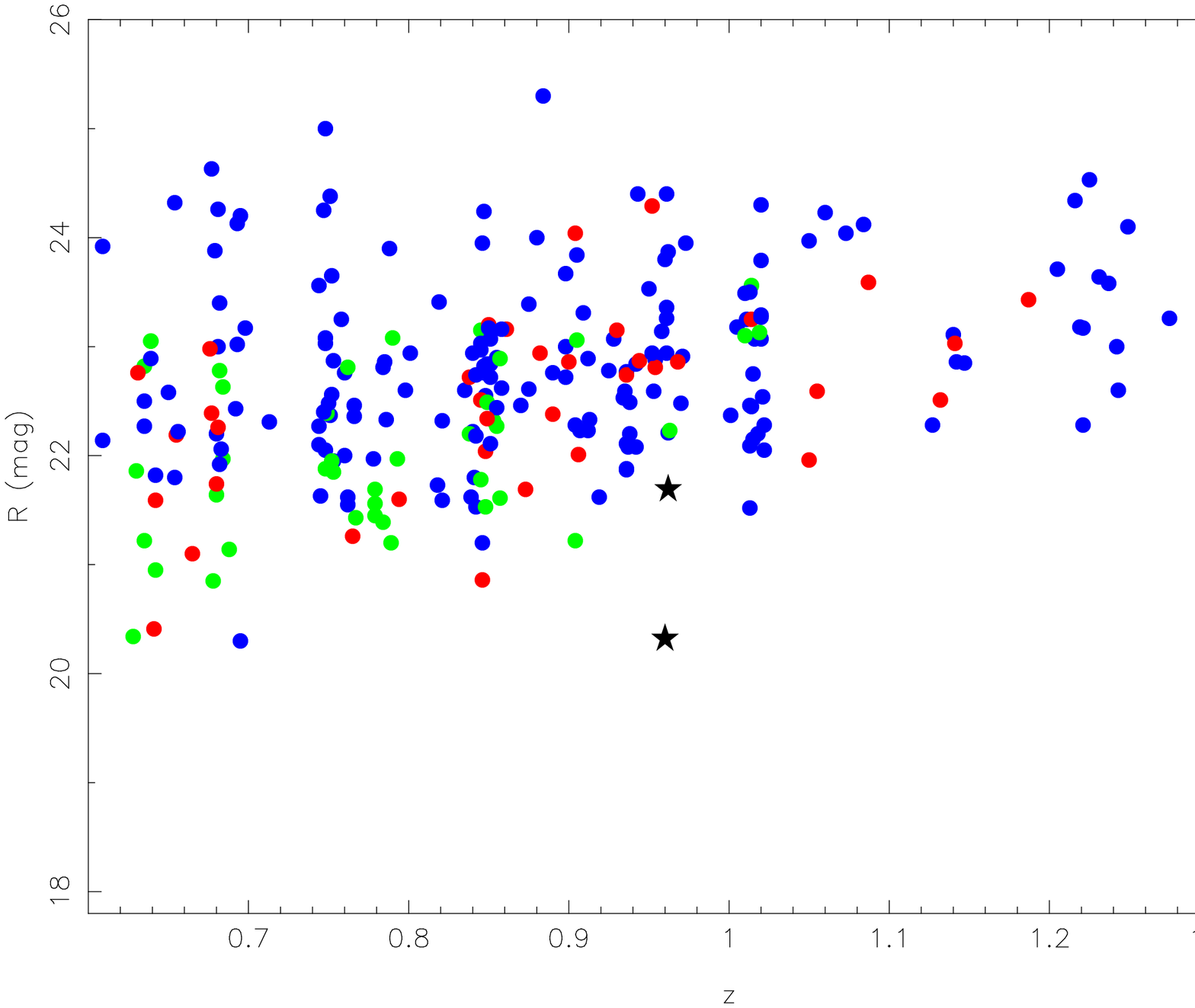}
\caption[figure8.ps]{b) The $R$ mag is shown as a function of redshift
for galaxies in the region of the HDF for $z < 0.7$ (figure 8a) and for
$0.6 < z < 1.4$ (figure 8b).  The symbols are the same as those of Figure~6.
\label{fig8}}
\end{figure}

%
\begin{figure}
\epsscale{0.9}
\plotone{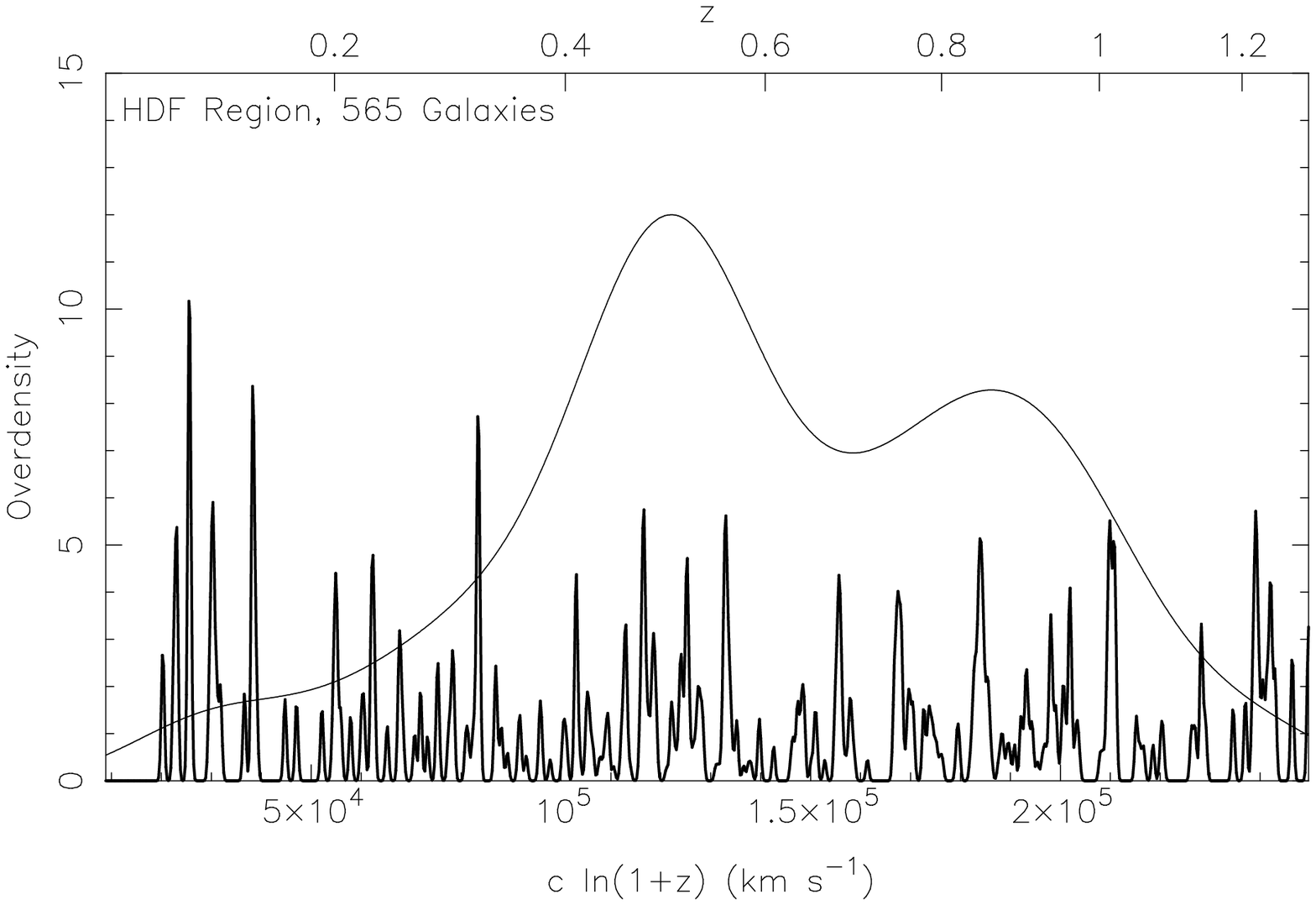}
\caption[figure9.ps]{The thick curve shows the overdensity as a function
of the local velocity, while the thin curve denotes the heavily smoothed
distribution of galaxies scaled by a constant.
The statistically significant sample
of redshift peaks consists of those with maximum overdensity
greater than 3. \label{fig9}}
\end{figure}

\begin{figure}
\epsscale{0.8}
\plotone{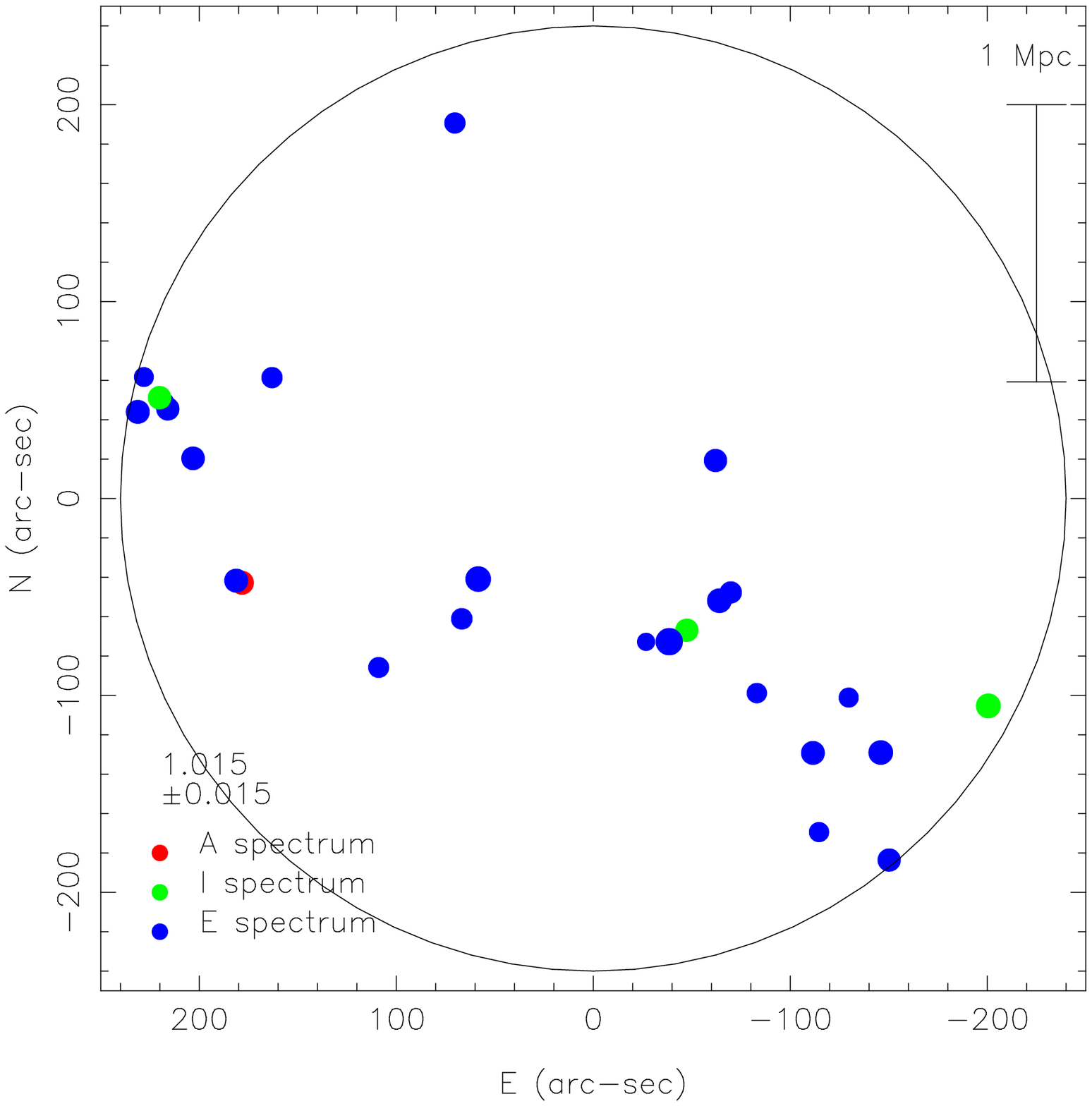}
\caption[figure10.ps]{The distribution on the sky of the galaxies with 
$z = 1.015 \pm0.015$ in the region of the HDF.  
The color coding indicates the galaxy spectral type as
in earlier figures.  The size of the circle increases with the apparent
brightness at $R$ of the galaxy.
\label{fig10}}
\end{figure}

\begin{figure}
\epsscale{0.8}
\plotone{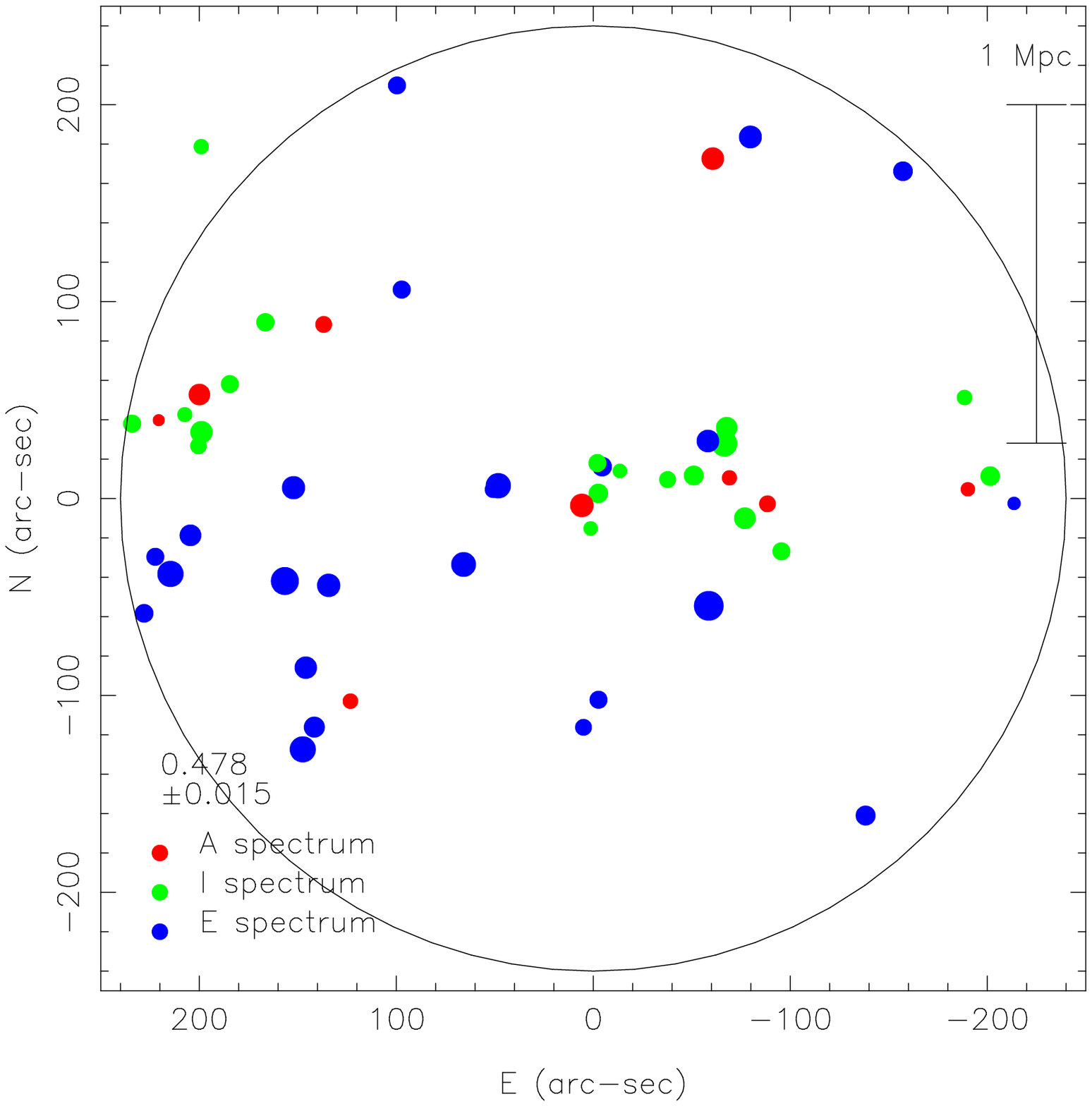}
\caption[figure11.ps]{The  distribution on the sky of the galaxies with 
$z = 0.478 \pm0.015$ in the region of the HDF.  The symbols are identical
to those of figure~10.
\label{fig11}}
\end{figure}

\clearpage

\begin{figure}
\epsscale{0.7}
\plotone{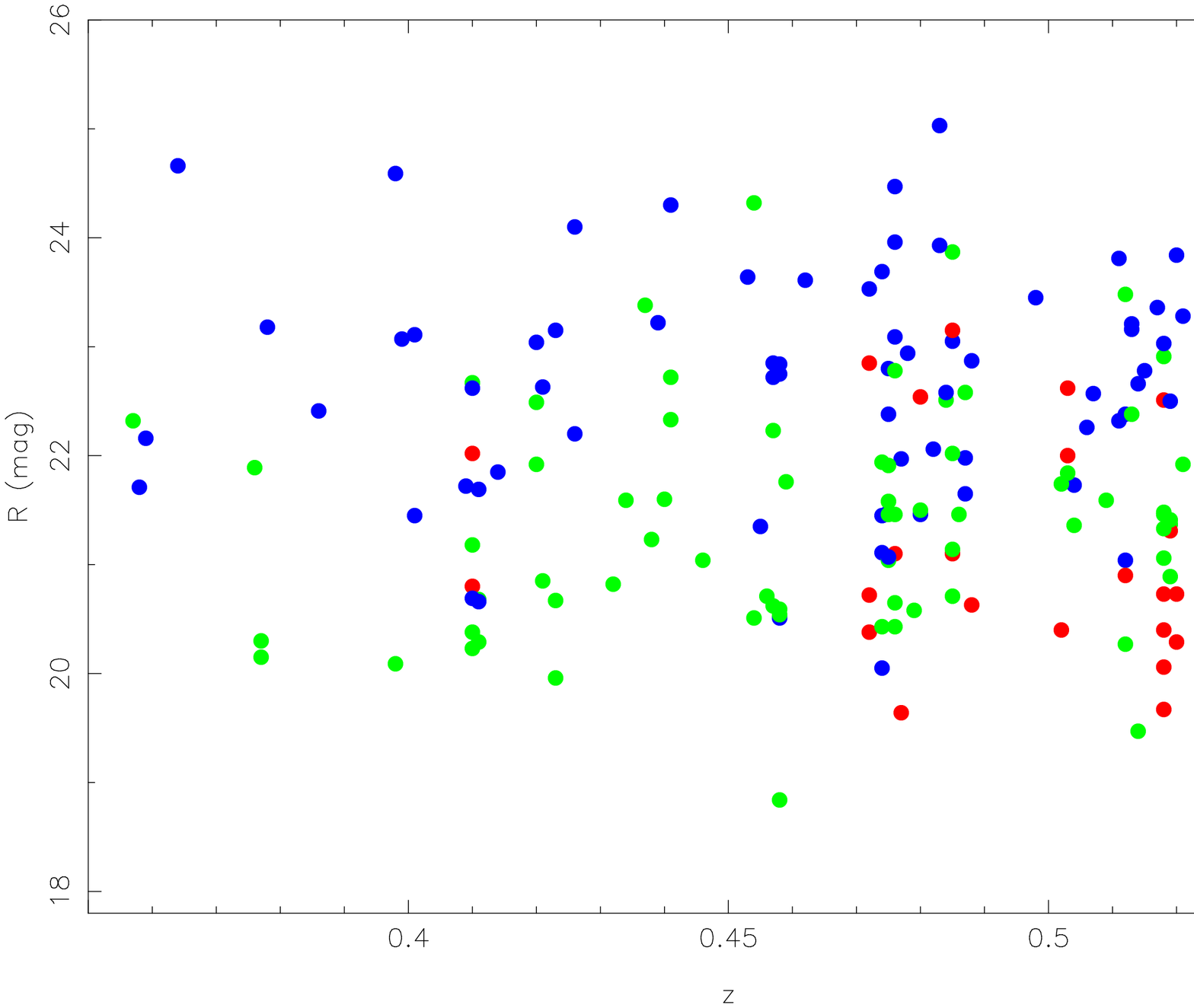}
\plottwo{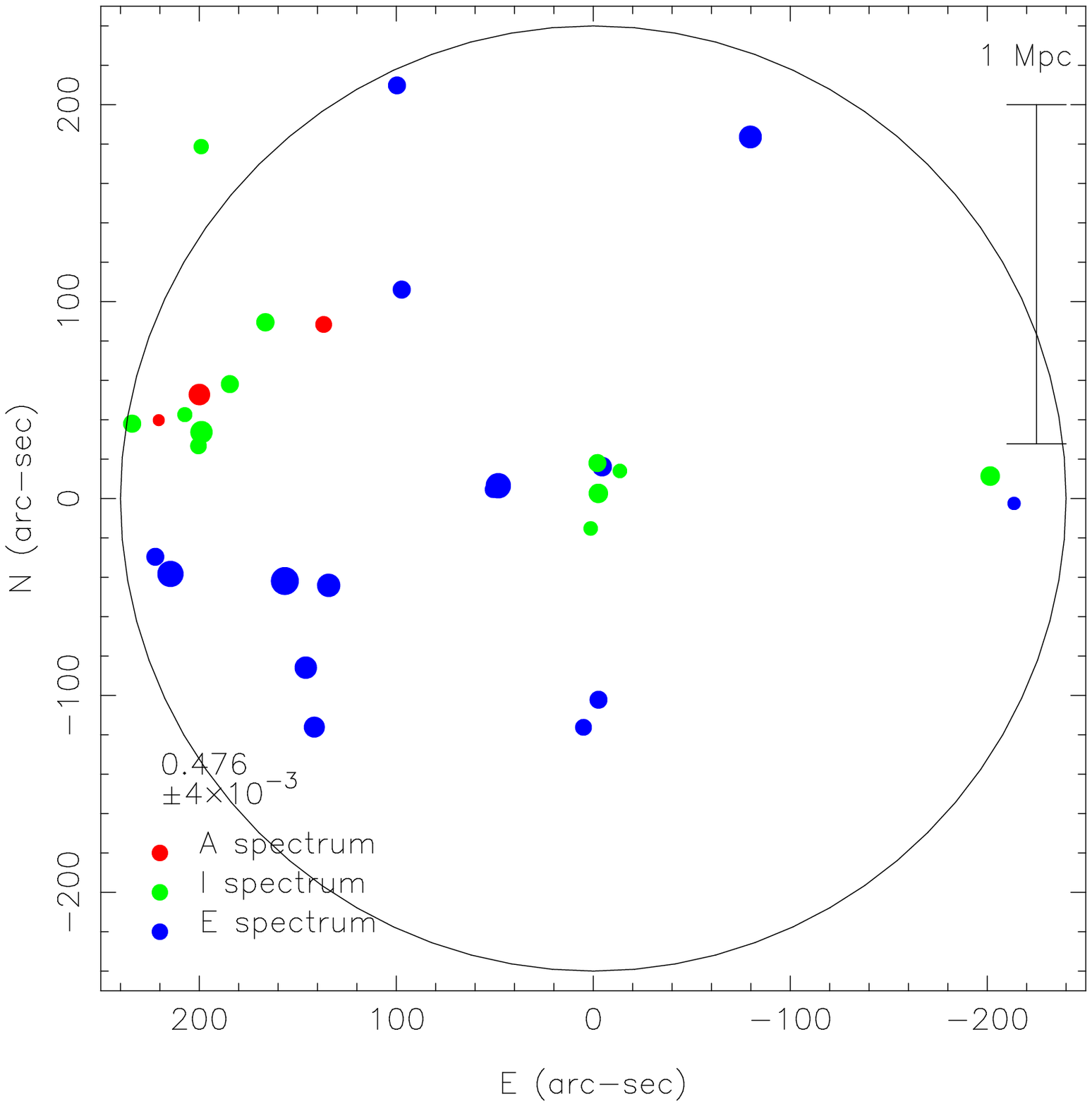}{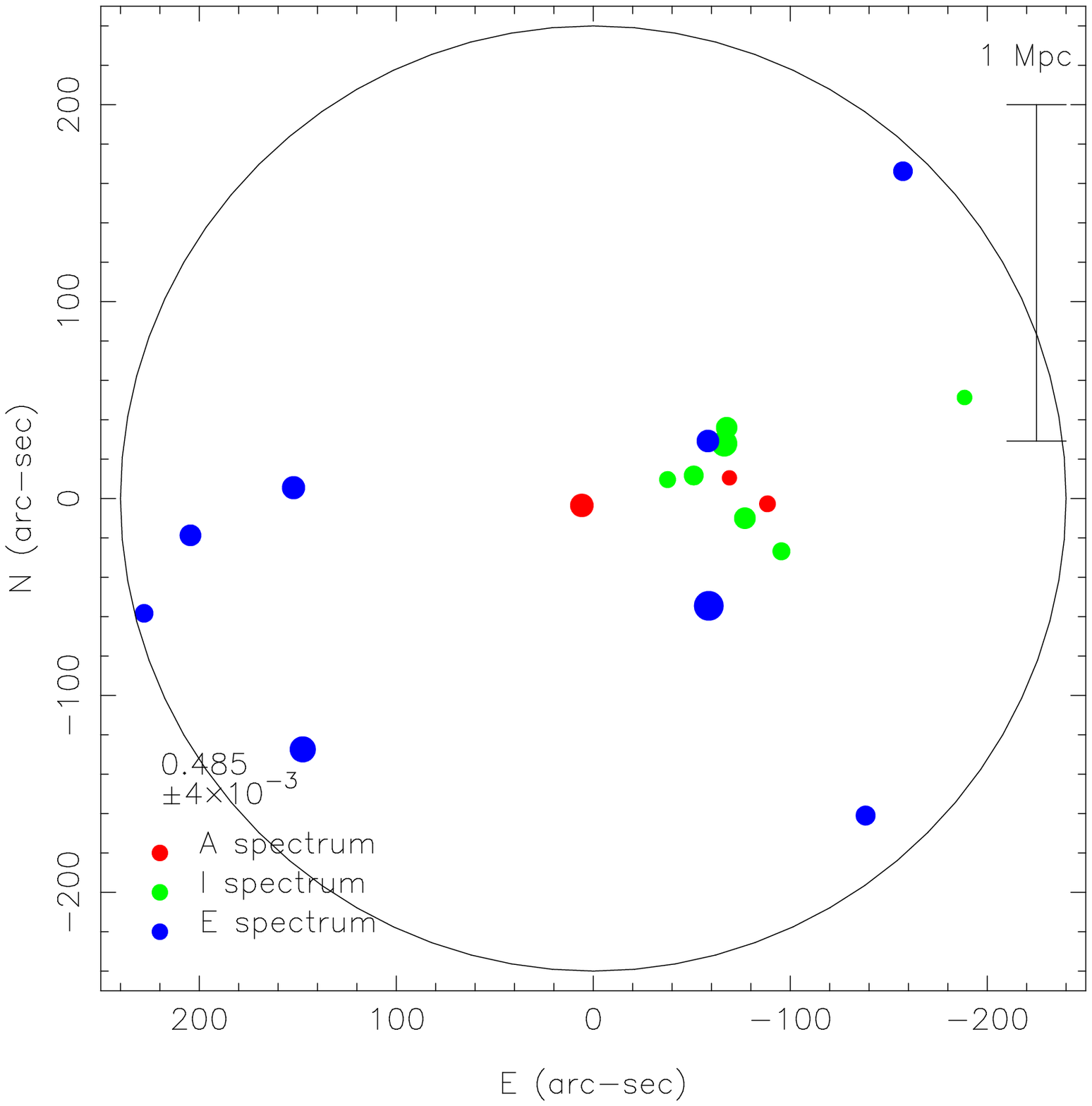}
\caption[figure12.ps] {The first panel shows the region $z \sim 0.5$
of the $z$ distribution 
of Figure~8a in more detail.  The second and  third panel show the spatial
distribution of the galaxies in the two groups that together comprise the
redshift peak at $z_p = 0.478$.
The symbols and colors used to denote galaxy spectral types are the same as in Figure~6.
\label{fig12}}
\end{figure}

\clearpage

%
\begin{figure}
\epsscale{0.8}
\plotone{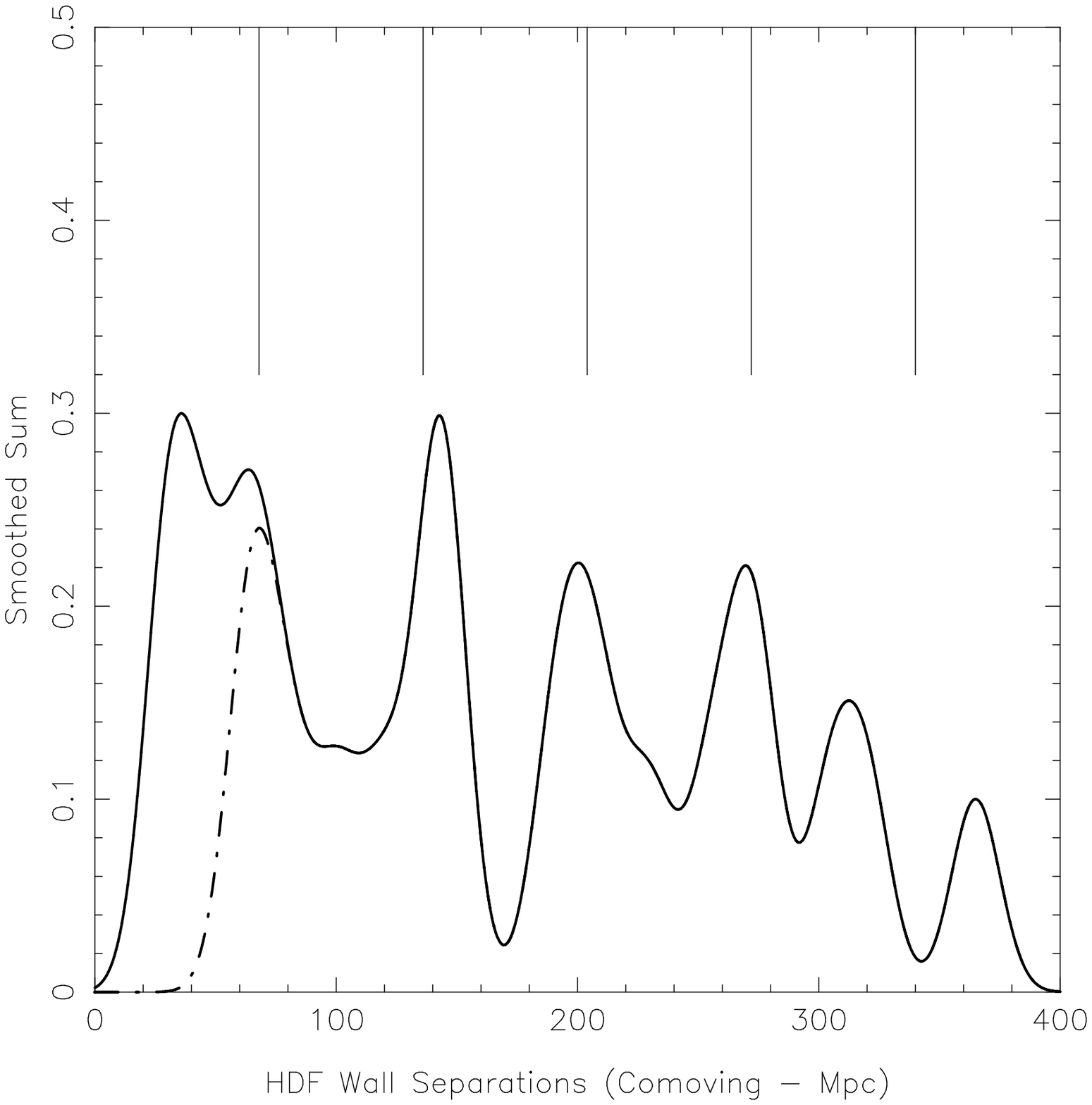}
\caption[figure13.ps] {The
histogram of the separation of adjacent redshift peaks from the
statistically complete sample of Table~6 is shown smoothed
with a Gaussian with $\sigma = 10$ Mpc.  The dashed line represents
the distribution when those separations under 50 Mpc,
which presumably are group-group separations within a single
redshift ``wall'', are omitted.  The vertical lines indicate
a period of 68 Mpc.
\label{fig13}}
\end{figure}

\begin{figure}
\epsscale{0.7}
\plotone{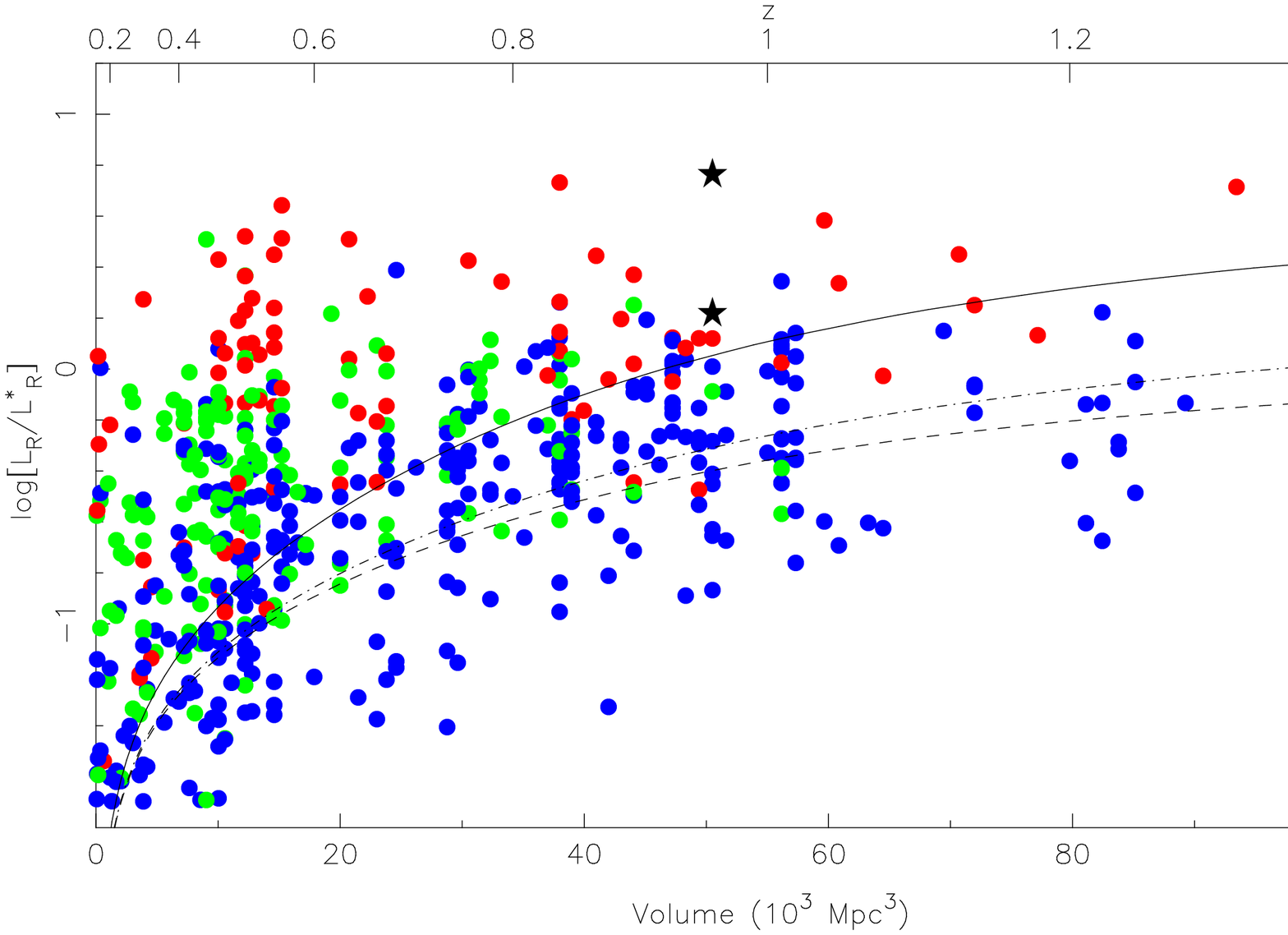}
\caption[figure14.ps] {The luminosity $L_R$ in units of 
$L_R^{\ast}$ is shown as a function of cosmological volume.
The corresponding redshift is shown at the top, while at
the right $L_R$ is given in units of W.  
The symbols and colors used to denote galaxy spectral types are the same as in Figure~6.
The lines denote the
survey cutoff for the HDF Flanking Fields at an apparent magnitude of $R= 23$ 
for elliptical (solid line), Sa 
and Sc galaxies using the passive evolution models of Poggianti (1997).
\label{fig14}}
\end{figure}

\begin{figure}
\epsscale{0.7}
\plotone{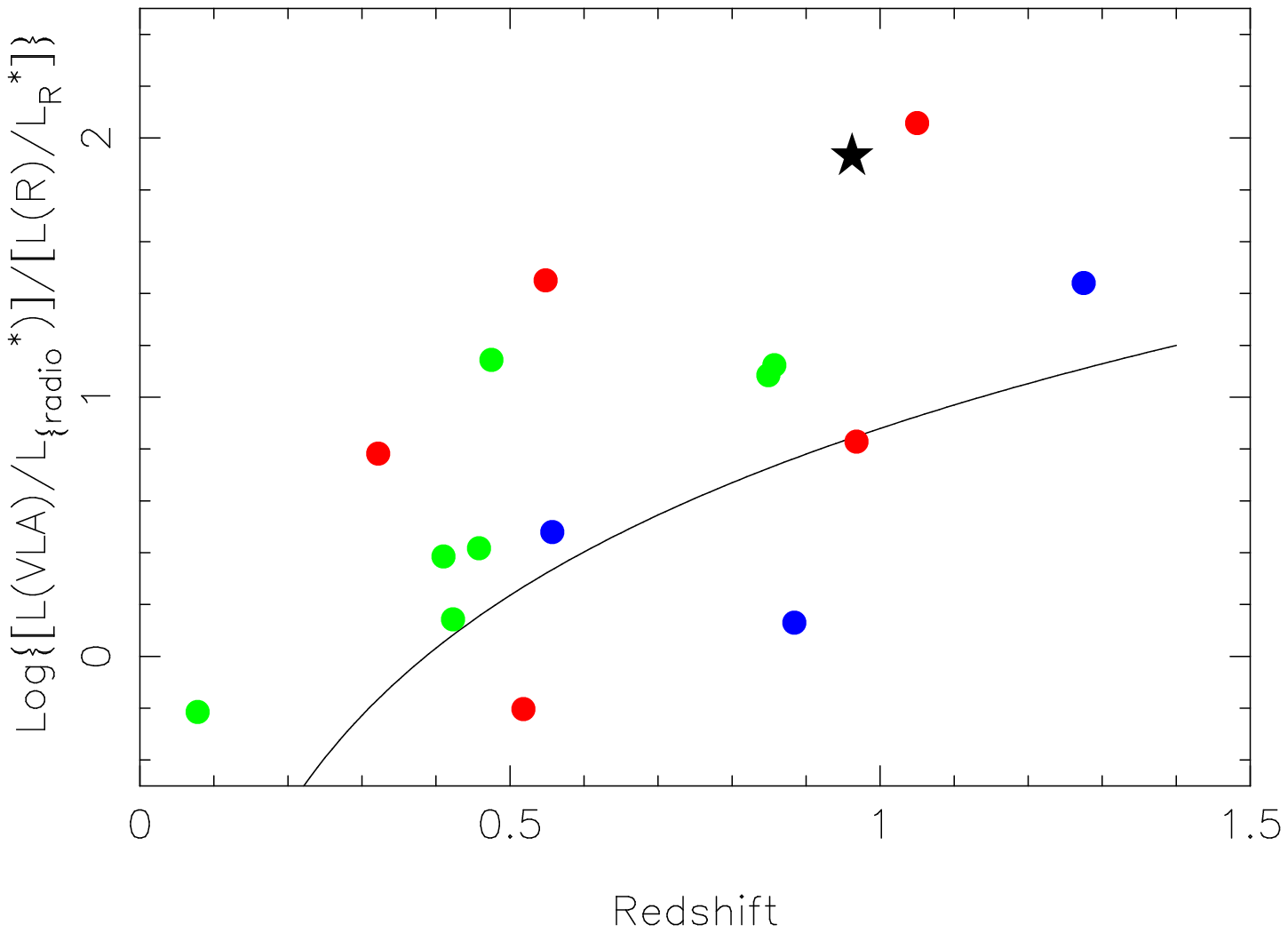}
\caption[figure15.ps] {The ratio of the VLA to the optical luminosity is shown
as a function of $z$
for the sample of VLA sources of Richards \etal\ (1998) with peak flux at 8.5 GHz
exceeding 9 $\mu$Jy for which
a secure optical counterpart with a redshift exists.  The optical luminosity is
rest frame $R$ in units of $L_R^*$.  The VLA luminosity is in units of $L_{8.5GHz}^*$
for local spiral and irregular galaxies with a $K$-correction based on mean spectral
index.
The symbols and colors used to denote galaxy spectral types are the same as in Figure~6.
The curve represents the selection limit imposed by a galaxy with $L = L_R^*$
and the minimum flux at 8.5 GHz for a secure detection.
\label{fig15}}
\end{figure}

\begin{figure}
\epsscale{0.7}
\plotone{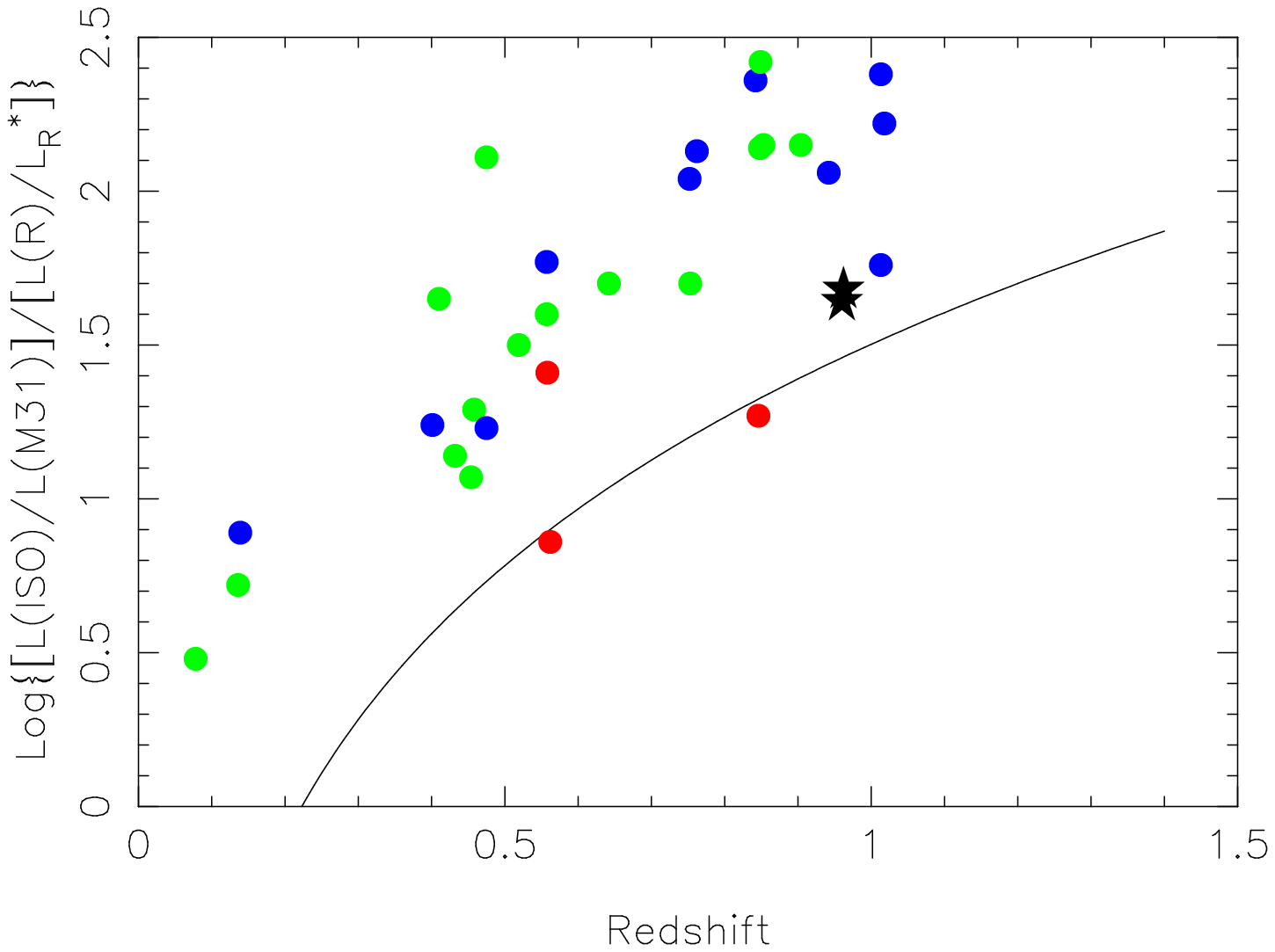}
\caption[figure16.ps] {The ratio of the ISO to the optical luminosity is shown
as a function of $z$
for the sample of highly significant ISO sources of Aussel \etal\ (1999) for which
a secure optical identification and a redshift exist.  The optical luminosity is
rest frame $R$ in units of $L_R^*$.  The ISO luminosity is in units of the
flux of M31 at 12$\mu$. 
The symbols and colors used to denote galaxy spectral types are the same as in Figure~6.
The curve represents the selection limit imposed by a galaxy with $L = L_R^*$
and the minimum flux for a secure detection with ISO, taken to be 40 $\mu$Jy.
No $K$-correction has been applied to the ISO data.
\label{fig16}}
\end{figure}

\end{document}